\documentclass[aps,pre,reprint,twocolumn,superscriptaddress,showpacs, longbibliography]{revtex4-1}
\usepackage{amssymb,amsmath}
\usepackage[table]{xcolor}
\usepackage{graphicx}
\usepackage{mathtools}
\usepackage[export]{adjustbox}
\usepackage{overpic}
\usepackage[colorlinks=true,
 linkcolor=black,
 urlcolor=blue,
 citecolor=blue]{hyperref}
\usepackage{nicefrac}
\usepackage{stackrel}
\usepackage{multirow}
\usepackage{lipsum} 
\usepackage[normalem]{ulem}
\usepackage{pbox}
\newcolumntype{C}[1]{>{\centering\let\newline\\\arraybackslash\hspace{0pt}}m{#1}}

\usepackage{enumitem}
\usepackage{bbold}

\usepackage{framed,color}
\definecolor{shadecolor}{rgb}{0.85,0.80,0.80}

\definecolor{myorange}{RGB}{253, 184, 99}
\definecolor{mypurple}{RGB}{178, 171, 210}

\newcommand{\comments}[1]{}
\usepackage{multirow}

\usepackage{float}

\newcommand{\beq}{\begin{equation}}
\newcommand{\eeq}{\end{equation}}
\newcommand{\bal}{\begin{aligned}}
\newcommand{\eal}{\end{aligned}}

\newcommand{\be}{\begin{equation}}
\newcommand{\ee}{\end{equation}}
\newcommand{\bd}{\begin{displaymath}}
\newcommand{\ed}{\end{displaymath}}
\newcommand{\BE}{\begin{eqnarray}}
\newcommand{\EE}{\end{eqnarray}}

\allowdisplaybreaks
\begin{document}
\title{Intrinsic noise, Delta-Notch signalling and delayed reactions promote sustained, coherent, synchronised oscillations in the presomitic mesoderm }
\author{Joseph W. Baron}
\email{joseph.baron@postgrad.manchester.ac.uk}
\affiliation{Theoretical Physics, School of Physics and Astronomy, The University of Manchester, Manchester M13 9PL, United Kingdom}

\author{Tobias Galla}
\email{tobias.galla@manchester.ac.uk}
\affiliation{Theoretical Physics, School of Physics and Astronomy, The University of Manchester, Manchester M13 9PL, United Kingdom}
\affiliation{Instituto de F{\' i}sica Interdisciplinar y Sistemas Complejos IFISC (CSIC-UIB), 07122 Palma de Mallorca, Spain}
 
\begin{abstract}
Using a stochastic individual-based modelling approach, we examine the role that Delta-Notch signalling plays in the regulation of a robust and reliable somite segmentation clock. We find that not only can Delta-Notch signalling synchronise noisy cycles of gene expression in adjacent cells in the presomitic mesoderm (as is known), but it can also amplify and increase the coherence of these cycles. We examine some of the shortcomings of deterministic approaches to modelling these cycles and demonstrate how intrinsic noise can play an active role in promoting sustained oscillations, giving rise to noise-induced quasi-cycles. Finally, we explore how translational/transcriptional delays can result in the cycles in neighbouring cells oscillating in anti-phase and we study how this effect relates to the propagation of noise-induced stochastic waves. 
\end{abstract}

\maketitle

\section{Introduction}

In developing vertebrates and cephalochordates, as the embryo forms and extends pairs of blocks of mesodermal progenitor cells assemble, bilaterally flanking the notochord \cite{pourquie}. These blocks, termed somites, eventually go on to form vertebrae and ribs after further cellular differentiation. The somites are constructed pair-by-pair, anterior to posterior, in a rythmic and sequential manner as the tailbud extends away from the rostral end of the embryo. They are formed from cells originating in the presomitic mesoderm (PSM). Such cells are produced continually by the tailbud as the abdomen elongates \cite{oatesreview, takedareview}. 

The process of somite segmentation has been of interest to experimentalists and theorists working in the field of developmental biology for some decades; it provides a fascinating case study where one can directly examine the link between microscopic gene regulatory systems operating in individual cells and macroscopic developmental processes. The prevailing theoretical framework for understanding the process was put forward by Cooke and Zeeman in 1976 \cite{cookezeeman} and is termed the `clock-wavefront' model. This model proposes that the cells in the presomitic mesoderm each possess an internal cyclic `clock' which is synchronised between the cells. Additionally, a wavefront propagates through the PSM as the embryo grows. As the wavefront encounters cells, it interacts with them differently depending on the current state of each internal cellular clock. This interaction causes the cells to change their adhesive and migratory properties. The temporal periodicity of the cell cycles is thus converted into the spatial periodicity of the somites.

Considerable experimental and theoretical effort has been expended in order to identify the genetic oscillators that constitute the putative somite segmentation `clock' and a good amount of progress has been made. In certain model organisms, such as the mouse and the zebrafish, so-called `knockdown'/`knockout' experiments have identified genes which when mutated give rise to defects in the formation of somites and, consequently, the vertebrae \cite{bessho, henry,oateshairy,sieger,conlon,deangelis,kusumi}. Gradients of FGF (fibroblast growth factor) or Wnt protein, which are produced in the tailbud, are thought to constitute the moving wavefront; transient loss or increase in these substances can alter the local somite length \cite{aulehla, dubulle}. Genes such as \textit{hes} in the mouse \cite{jouve} and \textit{her} in the zebrafish \cite{holley} are thought to be the primary cyclic genes which act as clocks. These genes are both targets of the Notch signalling pathway. It has also been shown in experiments that Delta-Notch signalling is a vital component in synchronising oscillations \cite{jouve,jiangnotch,ozbudaklewis, delaune}. 

In order for the oscillations in the expression of the \textit{hes/her} genes to constitute a viable segmentation clock for the clock-wavefront model, the oscillations must satisfy several criteria: (1) The oscillations in gene expression must have the same frequency in adjacent cells. (2) The oscillations in adjacent cells must be in phase. (3) The cellular oscillations must be coherent (there must be a clear dominant frequency). (4) The oscillations must have sizeable enough an amplitude so as not be indistinguishable from background `noise'. Mathematical models of the gene regulatory system have shown that Delta-Notch signalling can indeed synchronise (align the frequencies) of oscillations in neighbouring cells with intrinsically differing cellular clocks \cite{lewis}. That is, it has been shown that Delta-Notch signalling is responsible for satisfying condition (1), but relatively little discussion has been dedicated to the latter 3 conditions (phase, coherence and amplitude). 

So that one might analyse the degree to which the criteria above are satisfied, one must take into account stochastic (random) effects in the system, especially with regards to point (3). The gene regulatory systems in question are inherently noisy in nature  \cite{ozbudak2002, austin,shimojo, papalopulu1, papalopulu2, hirata, shimojoneuron}. This is due in part to the stochastic nature of the production/decay events of individual proteins and/or mRNA molecules and the fact that there are finite numbers of these molecules in any one cell. Noise of this origin is termed \textit{intrinsic} in the literature \cite{elowitz,elowitz2}. On the other hand, gene expression is also influenced by the concentrations, locations and states of molecules such as regulatory proteins or polymerases which can affect the global activity in a single cell but can vary between cells. Noise arising from fluctuations in the properties of such molecules is referred to as  \textit{extrinsic} \cite{elowitz,elowitz2}. 

The role of noise has largely been disregarded in previous theoretical work on the somite segmentation clock \cite{jensen,goldbeter,momijimonk}, or has often been treated only as an external influence rather than as an aspect intrinsic to the translation and transcription processes \cite{cinquin, morelli, horikawa} (an exception can be found in \cite{ay}, where simulations involving intrinsic noise were performed). For example, some works have considered the binding and unbinding of repressor protein to the DNA binding site as a stochastic process \cite{lewis,bakerreview} so that the rates of transcription are themselves stochastic variables. This source of noise is taken into account in the context of deterministic evolution equations - the intrinsic stochasticity of the transcription and translation events is not accounted for. In this paper however, we study the effect of intrinsic stochasticity by treating the production and the decay of individual molecules as random processes, following \cite{ozbudak2002, galla2009, barrio, schlicht, turnerstochastic, thattai}. These events may be subject to delays arising from the finite time taken for the translation/transcription processes. 

Using an individual-based mathematical model capturing the intrinsic noise in the system, we demonstrate that, perhaps counter to intuition, intrinsic stochasticity can be a proactive force in promoting cellular oscillations. We study how these noisy oscillations in neighbouring cells are affected by different levels of the strength of Delta-Notch signalling. We are able to show that, under certain conditions, Delta-Notch signalling not only acts to align the frequencies of oscillations in neighbouring cells; it can also reduce the phase lag, reduce the range of dominant oscillatory frequencies (i.e., it can make the oscillations more coherent) and it can increase the amplitude of oscillations (also noted in \cite{ay, ozbudaklewis}). The combination of these effects indicates that Delta-Notch signalling can contribute to satisfying points (1)-(4) above. We also discuss circumstances under which pairs of cells may oscillate out-of-phase, despite Delta-Notch coupling. We explore how this is related to waves and/or oscillating chequerboard patterns of gene expression in extended chains of cells.

\section{Methods}
\subsection{Model definition}\label{section:model}

Oscillations in the expression of genes (or `pulsing dynamics') is a well-documented phenomenon responsible for many cell functions and broader biological processes \cite{levinereview, goldbeterbook}. Such oscillations in the expression of certain genes are thought to constitute the biological `clock' in the clock-wavefront model \cite{cookezeeman} of somite segmentation \cite{takedareview,oatesreview}. The genes in question are known to be affected by Notch signalling. For the purposes of our theoretical treatment it is not necessary to consider the full complexity of the Notch signalling pathway \cite{brayreview, andersson} or even the full network of interacting genes involved with the somite segmentation process \cite{pourquie2}; one can use a simplified model to highlight the salient features and analyse their causes.

So-called `knockdown'/`knockout' experiments \cite{takedareview} suggest that the most relevant genes for the regulation of the `clock' are \textit{delta} (or its homologues) and \textit{hes} in mice, and \textit{her} in zebrafish \cite{oatesreview}. It has been shown previously \cite{momijimonk} that a two-gene model involving only \textit{hes/her} and \textit{delta} is sufficient for the emergence of cycles. We therefore also adopt a reduced two-gene model, as this will be sufficient to highlight the effects with which we are concerned. The reduced system is depicted schematically in Fig. \ref{fig:schematic} and is discussed in more detail in the Supplement (Section S1). For now, we consider a system of two coupled cells as a simple example. We generalise the approach to systems of greater numbers of cells in Section \ref{section:waves}.

\begin{figure}[H]
	\centering
    \includegraphics[scale = 0.22]{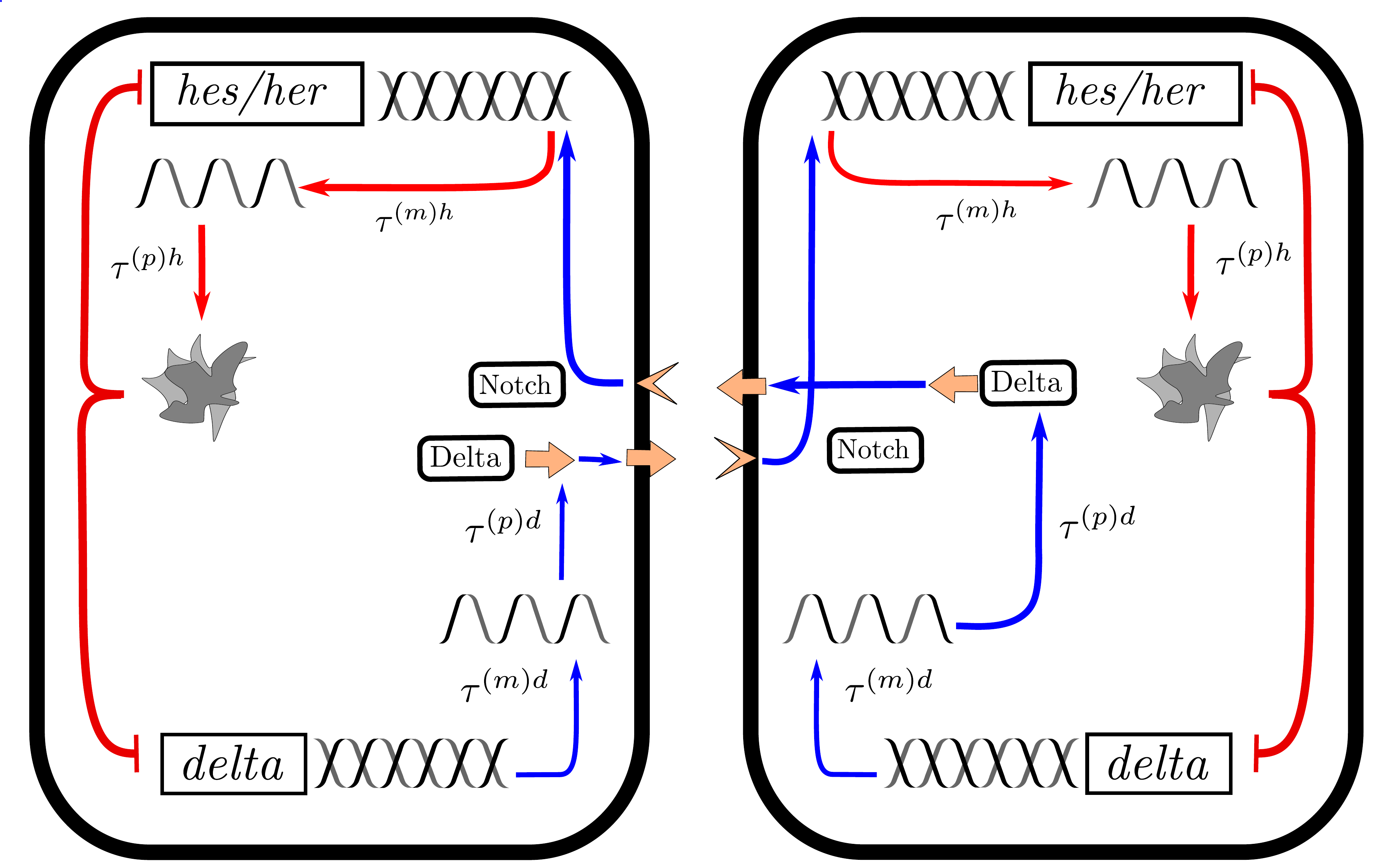}
	\caption{Schematic of the reduced 2-cell gene regulatory system \cite{lewis, momijimonk}. Genes (\textit{hes/her} and \textit{delta}) are transcribed to produce mRNA. In turn, protein is translated from the mRNA, which goes on to activate/inhibit further mRNA transcription. Both the transcription and translation processes take an amount of time to complete, giving rise to delays $\tau^{(m)x}$ and $\tau^{(p)x}$ respectively, where $x \in \{h,d\}$. Hes/Her protein inhibits local \textit{delta} transcription. Delta protein acts as a ligand to the Notch receptor on the adjacent cell. Notch in turn activates the production of \textit{hes/her} mRNA. Individual mRNA and protein molecules degrade (and become inactive) at constant probabilities per unit time. }
	\label{fig:schematic}
\end{figure}

Using $\mathbf{n}\left(t\right)$ to denote the set of all protein and mRNA numbers in all cells at time $t$, the reduced gene regulatory system that we consider can be summarised as follows: \textit{hes/her} mRNA is transcribed at a rate $f^h\left[\mathbf{n}\left(t- \tau^{(m)h}\right)\right]$. The rate $f^h\left[\mathbf{n}\left(t- \tau^{(m)h}\right)\right]$ takes the form of a sum of Hill functions, which reflects the fact that \textit{hes/her} mRNA production is inhibited by local Hes/Her protein and activated by Delta protein in adjacent cells. The precise form of this non-linear function is given in the Supplement (Section S1). Every \textit{hes/her} mRNA molecule is translated into protein at a constant rate $a_h$. Because the transcription and translation processes take an amount of time $\tau^{(m)h}$ and $\tau^{(p)h}$ to complete respectively, the present rate of production of mRNA/protein is dependent upon protein/mRNA concentrations in the past (respectively). Both \textit{hes/her} mRNA and protein molecules degrade (become inert) at constant per capita rates $c_h$ and $b_h$ respectively. In a similar way, \textit{delta} mRNA is transcribed at a rate $f^d\left[\mathbf{n}\left(t- \tau^{(m)d}\right)\right]$ and decays at a constant per capita rate $c_d$. Delta protein molecules are produced and decay at the per capita rates $a_d$ and $b_d$ respectively. Production of \textit{delta} mRNA and protein are delayed by the times $\tau^{(m)d}$ and $\tau^{(p)d}$ to respectively. Finally, production of \textit{delta} mRNA is inhibited by local Hes/Her protein concentration.

There are three vital aspects to the processes in this setup: (1) The model is individual-based -- it does not treat protein/mRNA concentrations as continuous quantities (an approximation only valid when population numbers are large). The production and degradation of proteins and mRNA are inherently stochastic (random) processes due to the finite numbers of proteins and mRNA \cite{papalopulu1, papalopulu2} in each cell; this gives rise to noisy dynamics \cite{turnerstochastic}. (2) There is a time-delay between the activation of the production of one unit of mRNA/protein and the completion of the production process. As a result, the rates of production of mRNA/protein at a given time are dependent on the state of the system in the past. In the language of stochastic processes, the dynamics are non-Markovian (they have memory) \cite{gardiner}. It has been established that time-delays such as these can encourage the emergence of temporal oscillations \cite{momijimonk, lewis, smithdelay}. (3) Due to Delta-Notch signalling, the rate of production of \textit{hes/her} mRNA in one cell is dependent on the concentration of Delta protein in the neighbouring cells. In this sense, there is a non-locality to the reaction rates.

The combination of these three aspects of the dynamics leads to a unique challenge with respect to theoretical modelling. However, we demonstrate in the Supplement that one can approximate the full individual-based dynamics of the system with a set of stochastic differential equations (SDEs); these are given in the Supplement (Section S2 C). They take a similar form to the deterministic (noiseless) equations given in \cite{lewis} but include additional Gaussian noise terms which take into account the intrinsic stochasticity of the system. We emphasize that the properties of this noise are calculated so as to agree with individual-based simulations of the system; the noise is not added in an \textit{ad hoc} fashion. The tools we use to quantify the phenomena induced in the system by noise are discussed in the next section.

\subsection{Analysis of stochastic behaviour}\label{section:stochasticbehaviour}
In this work, we will be concerned primarily with the theoretical analysis of noise-induced cycles of gene expression. These are oscillations which occur in the full stochastic individual-based model but which are missing in the noiseless deterministic system.

The power spectrum of fluctuations about the deterministic trajectory will be the main quantitative tool that we use to analyse the noise-induced phenomena in the gene regulatory model described in Section \ref{section:model} (and elaborated upon in the Supplement Section S1). We denote the number of particles of type $\alpha$ in cell $j$ at time $t$ by $n^\alpha_{j}(t)$. The type of particle indicated by the index $\alpha$ may be mRNA molecules or proteins. The dynamics of the quantities $n^\alpha_{j}(t)$ are approximated by the system of stochastic differential equations (SDEs) in Eq.~(S25) (in the Supplement). Further, we write $\bar n^\alpha_j\left(t\right)$ for the numbers of particles predicted by the corresponding deterministic model [Eq.~(S25), with the noise terms $\xi^\alpha_j\left(t\right)$ set to zero]. Thus, we define the fluctuations about the deterministic trajectory as 
\begin{align}
\delta^\alpha_j\left(t\right) = n^\alpha_j\left(t\right) - \bar n^\alpha_j\left(t\right). \label{deviations}
\end{align}
The power spectrum of fluctuations is then defined via the temporal Fourier transform as
\begin{align}
P_j^\alpha\left(\omega\right) = \langle \lvert\hat{\delta}^\alpha_j\left(\omega\right) \rvert^2 \rangle ,  \label{powerspectrumdef}
\end{align}
where the Fourier transform is given by $\hat g\left(\omega\right) =\frac{1}{\sqrt{2\pi}}\int_{-\infty}^\infty e^{i\omega t} g\left(t\right) dt$; the angular brackets denote the ensemble average over the set of all possible stochastic time courses of the system. Roughly speaking, the power spectrum of fluctuations decomposes a time series into its composite frequencies and quantifies the statistical contribution of a particular frequency to the series. A large, narrow, unique peak in the power spectrum indicates that the frequency at which the peak is located is the dominant frequency of the time series; a peak of this type centred on a non-zero frequency therefore characterises periodicity. If the deterministic trajectory $\bar{n}^\alpha_j\left(t\right)$ is non-oscillatory, then such a peak in the spectrum of the stochastic model indicates noise-induced oscillations. We note that in our analysis the power spectrum is always evaluated in the steady state, i.e. when all transient effects have decayed sufficiently so as to be negligible. 

Using a mathematical approach (see Supplement Sections S1, S2 and S3), we are able to predict the power spectrum of the fluctuations when the deterministic trajectory has reached a fixed point (i.e., when $\bar{n}^\alpha_{j}\left(t\right) =  \bar{n}^\alpha_{j}$ is constant at long times $t$). This theoretical approach gives us a way to identify, without performing time-consuming simulations, what the dominant frequency of noise-induced oscillations is and to what extent the other frequencies contribute. This analysis is only valid for noise-induced cycles, i.e. when the oscillations of the deterministic equations are transient. For the parameter regimes where there are persistent cycles in the deterministic equations, we perform a linear stability analysis in order to obtain quantities such as periods of oscillation and phase lags (see Supplement Section S3). 

It is also possible to quantify the phase lag between two sets of noise-induced cycles in coupled cells using our theoretical approach. Following \cite{challengermckane,rozhnovamckane}, we define the phase lag $\phi^{\alpha, \alpha'}_{j,j'}\left(\omega\right)$ associated with a particular frequency $\omega$ between species $\alpha$ in cell $j$ and species $\alpha'$ in cell $j'$ as
\begin{align}
\tan\left(\phi^{\alpha,\alpha'}_{j,j'}\left(\omega\right)\right) = \frac{\mathrm{Im}\left(\langle \hat\delta^\alpha_{j}\left(\omega\right)\hat\delta^{\alpha'\star}_{j'}\left(\omega\right)\rangle \right)}{\mathrm{Re}\left(\langle \hat\delta^\alpha_{j}\left(\omega\right)\hat\delta^{\alpha'\star}_{j'}\left(\omega\right)\rangle\right)} .\label{phaselag}
\end{align}
Phases differing by integer multiples of $2\pi$ are degenerate therefore, in this paper, we define the phase lag to be in the range $[-\pi, \pi)$. Notably, the phase lag as defined in Eq.~(\ref{phaselag}) is dependent on $\omega$. As mentioned above, a time series can be thought of as being comprised of a sum of cycles with different frequencies $\omega$. The quantity $\phi^{\alpha,\alpha'}_{j,j'}\left(\omega\right)$ is the phase lag between the constituent cycles of frequency $\omega$ in cells $j$ and $j'$. In our analysis, we may refer to \textit{the} phase lag of a cell $j$ (with respect to another cell), which we define as the phase lag at the peak frequency of the power spectrum of the cell in question, $\omega^{(j)}_{\textrm{max}}$. 

Furthermore, following \cite{alonsomckane}, we define the total amplification of fluctuations for particles of type $\alpha$ in cell $j$
\begin{align}
A^\alpha_j = \int_0^\infty P^\alpha_j\left(\omega\right) d\omega.
\end{align}
This quantity is proportional to the time-averaged squared displacement of the dynamics from the fixed point, i.e., to the variance of the stochastic time series.

Finally, again following \cite{alonsomckane}, we also define the coherence as the proportion of the power spectrum within a fixed range $\Delta \omega$ of the peak
\begin{align}
C^\alpha_j = \frac{1}{A^\alpha_j}\int_{\omega^{(j)}_{\mathrm{max}}-\frac{\Delta \omega}{2}}^{\omega^{(j)}_{\mathrm{max}}+\frac{\Delta \omega}{2}} P^\alpha_j\left(\omega\right) d\omega . \label{coherencedef}
\end{align}
The coherence $C^\alpha_j$ quantifies how sharply peaked the power spectrum is - i.e., how narrow the band of dominant frequencies is. It has a maximum value of $1$ and a minimum value of $0$. The choice of $\Delta\omega$ is largely immaterial provided $\Delta\omega$ is small compared to the peak frequency. 

\section{Results}
\subsection{Individual-based models capture noise-driven effects which are missed by deterministic models }\label{section:deterministicmodels}
\begin{figure}[H]
	\centering
        \includegraphics[scale = 0.14]{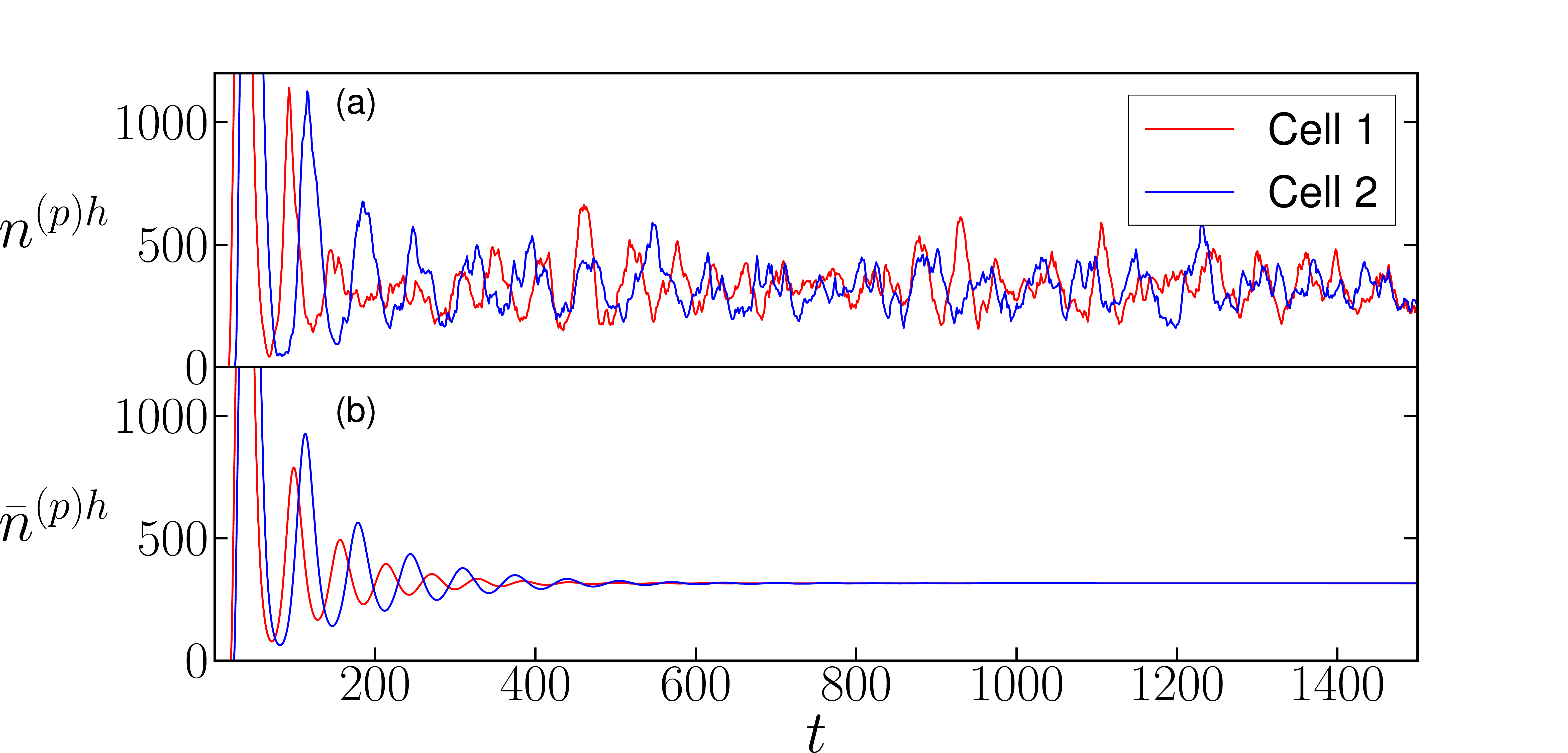}
	\caption{Oscillations in Hes/Her protein numbers for the simplified two-cell gene regulatory system (depicted in Fig. \ref{fig:schematic}) with no Delta-Notch coupling.  Panel (a) shows the results of individual-based simulations where noisy cycles persist. The simulations of the non-linear stochastic model are performed using a modified version \cite{nextreaction} of the Gillespie algorithm \cite{gillespie}, which takes into account delayed reactions. This is in contrast to the deterministic trajectory in panel (b), where the oscillations are transient and eventually decay to a fixed point. A small difference between the delays in either cell gives rise to differing frequencies of oscillation - the noisy cycles are not synchronised. This is further illustrated by the power spectra of the cycles in panel (a), which are shown in Fig. \ref{fig:spectraphase}(a2). Referring to the model specified in Supplement Section S1, the rate parameters used here are $a^\alpha = 4.5$, $b^\alpha = 0.23$, $c^\alpha = 0.23$ and $k_\alpha = 3.3$ for all $\alpha$, the system size is $N = 10$, the reference protein levels are $n^{\left(p\right)h}_0 = 4N$ and $n^{\left(p\right)d}_0 = 100N$, the delay times are $\tau^{\mathrm{\left(p\right)h}} = 2$, $\tau^{\mathrm{\left(p\right)d}} =5$ and  $\tau^{\mathrm{\left(m\right)d}} = 50$ in both cells but $\tau^{\mathrm{\left(m\right)h}} = 18$ in cell 1 and $\tau^{\mathrm{\left(m\right)h}} = 22$ in cell 2. The rates associated with the Hill functions are $r_0^{\mathrm{h}} =r_d^{\mathrm{h}} =r_{hd}^{\mathrm{h}}=  0$, $r_{h}^{\mathrm{h}}=1$, $r_0^{\mathrm{d}} =r_d^{\mathrm{d}} =r_{hd}^{\mathrm{d}}=  0$ and  $r_{h}^{\mathrm{d}}=1$. These values are taken from estimates provided in \cite{lewis}, which are justified therein. Times are in units of minutes, and rates have units of $\mbox{min}^{-1}$.}
	\label{fig:deterministicvsstochasticrh1point0}
\end{figure}

\begin{figure}[H]
	\centering
        \includegraphics[scale = 0.14]{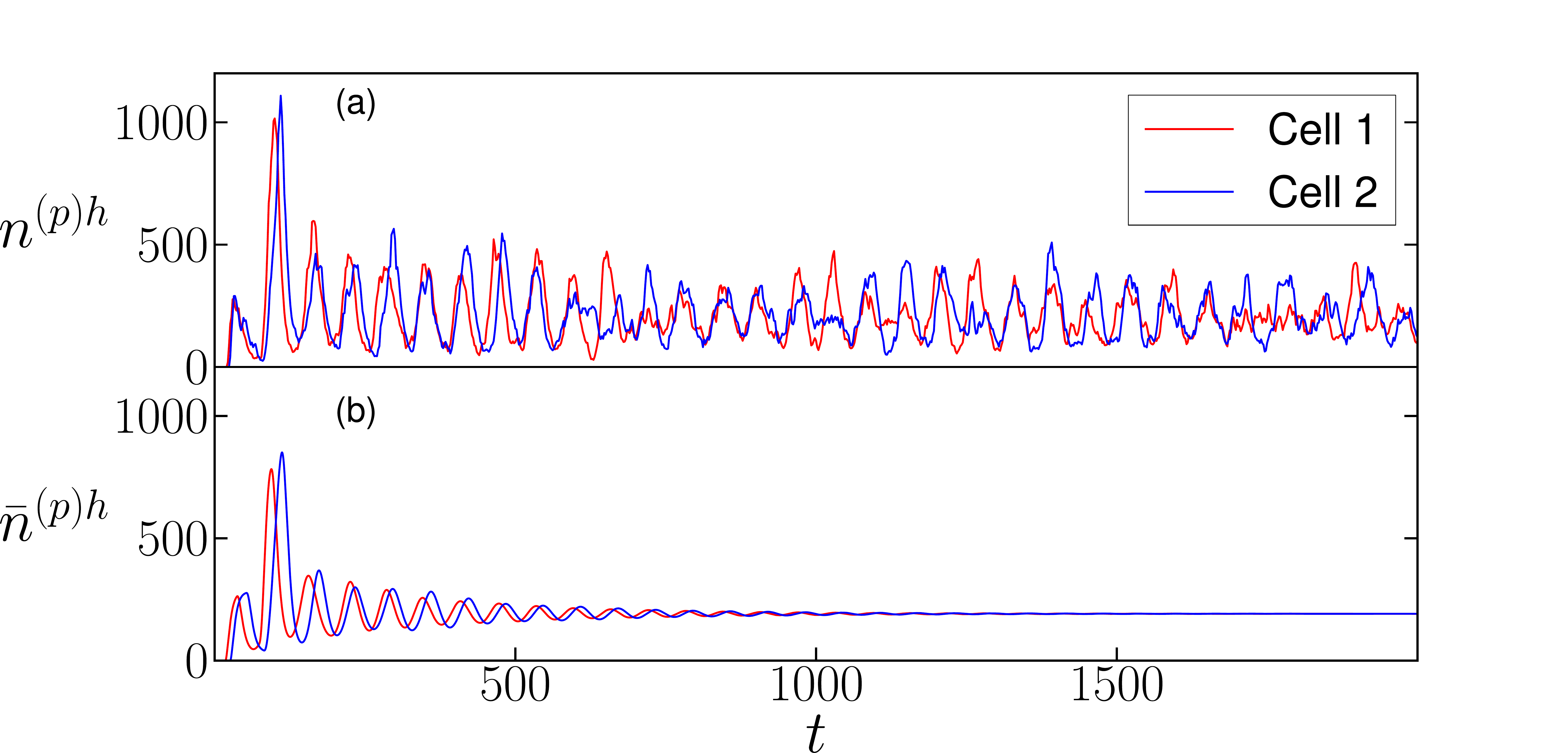}
	\caption{Oscillations in Hes/Her protein numbers for the simplified 2-cell gene regulatory system with Delta-Notch coupling. The system parameters are identical to Fig. \ref{fig:deterministicvsstochasticrh1point0}, but here $r_{hd}^{\mathrm{h}}=  0.9$ and $r_{h}^{\mathrm{h}}=0.1$, i.e. the coupling between the two cells has been increased (see Supplement Section S1). As a result, the oscillations in the individual-based system are more clear, periodic and synchronised. This is further demonstrated by the corresponding power spectra in Fig. \ref{fig:spectraphase}(c2). Again, the deterministic trajectory poorly reflects the dynamics of the individual-based system.  }
	\label{fig:deterministicvsstochasticrh0point1}
\end{figure}

In the systems we are considering, individual cells contain of the order of 10-100 mRNA molecules and around 1000 proteins of any one type \cite{papalopulu1, papalopulu2} (see also \cite{lewis, keskin}). As such, the dynamics are inherently noisy. This type of noisy dynamics has been observed in experiments monitoring gene expression \cite{austin, shimojo, papalopulu1, papalopulu2, hirata, shimojoneuron}. The expression of these genes cannot be fully described by the regular, smooth oscillation obtained from integrating deterministic sets of ODEs (as can be seen from Figs. \ref{fig:deterministicvsstochasticrh1point0} and \ref{fig:deterministicvsstochasticrh0point1}). Instead, a stochastic individual-based model is better suited to qualitatively reproduce the results of experiment.

In previous theoretical studies of the somite segmentation clock, noise has mostly been treated as external to the dynamics \cite{lewis, morelli, horikawa} or has not been considered at all \cite{momijimonk, kageyamareview}. An exception to this is \cite{ay}, in which individual-based simulations were carried out (the consequences of the inclusion of intrinsic noise that we discuss here were not the focus of \cite{ay} however). The noise that we use in the present work is rigorously derived as an intrinsic quality of the stochastic, individual-based dynamics themselves. As such, the results of our analysis agree with fully individual-based simulations of the system (as is demonstrated in Fig. \ref{fig:spectraphase}). 

It has previously been observed \cite{lewis} that noisy external driving can give rise to sustained oscillations in the somite segmentation clock. We show that similar noise-induced oscillations can also be produced by the stochastic nature of the transcription and translation processes in the segmentation clock themselves. That is, the inclusion of intrinsic noise in the theoretical modelling gives rise to sustained noise-induced oscillations which a purely deterministic model, or a more ad hoc approach to noise-inclusion, might miss. As is shown in Figs. \ref{fig:deterministicvsstochasticrh1point0} and \ref{fig:deterministicvsstochasticrh0point1},  for sets of parameters which are biologically reasonable (see \cite{lewis}), one may observe the noiseless model tend towards a stationary fixed--point, only exhibiting transient oscillations which eventually decay. In the corresponding individual-based model however, the noise repeatedly `kicks' the system away from the fixed point. As a consequence, the oscillations which were transient in the deterministic model are sustained by the noise. One thus observes persistent noisy oscillations for the same parameter set.

The oscillations in gene expression observed experimentally may very well be noise-induced cycles of this type. The emergence of such `quasi-cycles' is a well-documented phenomenon which has been previously studied in the context of gene-regulatory models \cite{galla2009, barrio} as well as in ecological systems \cite{lugomckane, mckanenewman} and in epidemics \cite{alonsomckane, rozhnovamckane, blackmckane}. That these are indeed cycles with a periodic nature and not just random white noise is demonstrated by the power spectra of fluctuations (see Fig. \ref{fig:spectraphase})-- this matter is discussed further in Sections \ref{section:stochasticbehaviour} and \ref{section:synchronisation}. 

Our theoretical approach to analysing noise-induced cycles (which is similar to that found in \cite{brettgalla1,brettgalla2,barongalla} and detailed in the Supplement) allows us to study the amplification, synchronisation and coherence of these cycles, as discussed in the following sections.

\subsection{Delta-Notch signalling mitigates inhomogeneity and promotes a robust and reliable segmentation clock in noisy oscillators}

Having introduced the concept of noise-induced cycles and the mathematical tools that we will use to analyse them, we now turn our attention to the effect that increasing the Delta-Notch signalling strength has on these noisy oscillations.

It has been observed experimentally \cite{jouve} that mutations in the \textit{delta} gene give rise to defects in the formation of somites. This has been attributed to a decreased coupling between the cells arising from the mutation which, due to slight inhomogeneities between cells and the stochastic nature of the cellular cycles, leads to the genetic oscillations in neighbouring cells becoming asynchronous \cite{jiangnotch, delaune}. Furthermore, it has been shown that encumbered Delta-Notch signalling (i.e. reduced signalling strength) can give rise to greater disparities between the oscillations in cells which would be synchronised if signalling were not impaired \cite{keskin}. In this section, we reproduce and study this effect with our model and theoretical approach (presented in the Supplemental material in Section S2), thus verifying the necessity of Delta-Notch signalling for the somite segmentation clock. A justification for our mathematical definition of `coupling strength' is given in Supplement Section S1.

Consider a two-cell system in which each cell has slightly different internal parameters (e.g. transcriptional delay time) such that the typical cycle time varies between the cells when they are uncoupled. We evaluate, using our theoretical approach, the response of the peak frequencies, inter-cell phase lag, amplification and coherence of genetic oscillations in this inhomogeneous two-cell system for various degrees of Delta-Notch coupling strength and thus show that the quality of the oscillations (and therefore the segmentation clock) can improve when Delta-Notch signalling is enhanced.

We use two different theoretical approaches in the following sections, each of which is valid for different, but complementary, parameter sets. (1) In the regime where the deterministic (noiseless) equations approach a fixed point, we evaluate the power spectrum of fluctuations of the emergent noise-induced cycles using the so-called linear-noise approximation (discussed in more detail Section S2 of the Supplemental Material). This analysis relies upon the deterministic dynamics tending towards a fixed point, and approximates the stochastic equations as linear in the vicinity of this fixed point. The accuracy of the approximation is tested against individual-based simulations of the full non-linear stochastic model in Fig. \ref{fig:spectraphase}. (2) When the fixed point of the deterministic system becomes unstable, we use a deterministic linear stability analysis (LSA) to find the dominant oscillatory frequency of the cycles and the inter-cell phase lag. The LSA provides no way of finding the amplitude or the coherence of the cycles however -- it is only possible for us to evaluate these when method (1) is valid.  

It has been suggested that a sufficiently strong non-linearity (the so-called `cooperativity') is required for regular sustained  cycles \cite{barrio}. Noting that our linear analysis agrees with simulation results (see Fig. \ref{fig:spectraphase}), we stress that the precise form of the non-linearity is not directly important for the emergence of noise-induced cycles. Their properties are well captured by the linearised dynamics. Non-linearities will however affect the location and nature of deterministic fixed points, and the coefficients in the linearised equations near these fixed points.

\subsubsection{Delta-notch signalling synchronises noisy genetic oscillators and reduces their phase difference}\label{section:synchronisation}

\begin{widetext}

\begin{figure}[H]
	\centering
        \includegraphics[scale = 0.25]{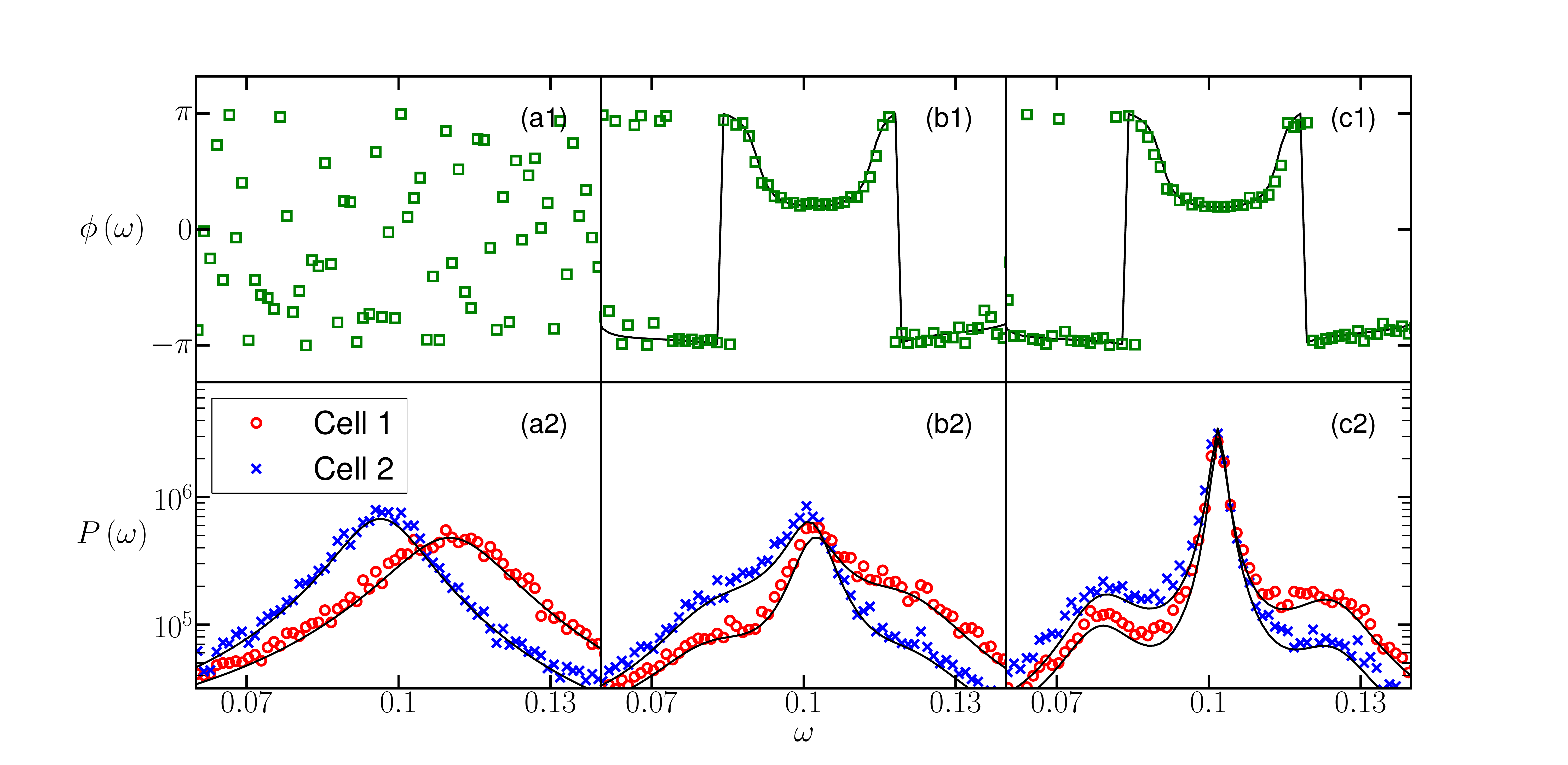}
	\caption{Synchronisation of stochastic oscillations of \textit{hes/her} expression in the 2-cell system as coupling strength is increased. The system parameters are as in Fig. \ref{fig:deterministicvsstochasticrh1point0} but with (a) $r^h_{hd} = 0$, (b) $r^h_{hd} = 0.7$ and (c) $r^h_{hd} = 0.9$, subject to the constraint $r^{\mathrm{h}}_{\mathrm{h}} + r^{\mathrm{h}}_{\mathrm{hd}} = 1 $. That is, the Delta-Notch coupling strength increases from (a) to (c). Panels (a1), (b1) and (c1) depict the phase lag $\phi^{(p)h}_{1,2}\left(\omega\right)$ between the oscillations of protein numbers in the two cells as a function of frequency $\omega$.  Panels (a2), (b2) and (c2) show the associated Fourier power spectra $P^{(p)h}_j(\omega)$ for both cells. In all panels, simulation results are represented by coloured markers whereas theory results are shown as black lines. The theory lines are produced using the analysis presented in Sections S2B and S2C of the Supplemental Material. Simulation results are averaged over 100 realisations of the system. In panel (a2), there is zero coupling and the peaks of the power spectra are separate, indicating different frequencies of oscillation in the two cells and a lack of synchronisation. In panel (a1), the phase lag between the two cells is random since the two cells oscillate independently. One observes that as the coupling strength is increased, the cells converge on a common frequency (i.e. they synchronise) and that this common frequency is one which minimises the phase lag between the cells.  The power spectra in panels (a2) and (c2) correspond to the time series in Figs. \ref{fig:deterministicvsstochasticrh1point0}(a) and \ref{fig:deterministicvsstochasticrh0point1}(a) respectively. }
	\label{fig:spectraphase}
\end{figure}

\end{widetext}

Firstly, we find that Delta-Notch signalling can have the effect of synchronising the dominant oscillatory frequencies of two cells with differing internal parameters. This is demonstrated in Fig. \ref{fig:spectraphase}, which depicts the power spectra of the stochastic fluctuations in two such cells for various degrees of coupling strength. One observes that as the coupling strength is increased, the peaks for either cell, which are separated when there is no coupling, are both drawn towards a common frequency, indicating synchronisation. The degree of synchronisation varies smoothly with the variation of the coupling strength, as shown in Fig. \ref{fig:variationwithcoupling}(a); in this figure, the dependence of the dominant frequencies on coupling strength is shown in more detail. The common frequency which is converged upon at large strengths of the Delta-Notch coupling agrees with that predicted by linear stability analysis (LSA) (see Supplement Section S3 and Section \ref{section:amplitudecoherence} for further details).

Secondly, we find that the peaks of the power spectra in either cell converge to a frequency which reduces the phase lag between the oscillations in the two cells (see Fig. \ref{fig:spectraphase}). So, not only can Delta-Notch signalling act to align the frequency of oscillations in neighbouring cells, it can also encourage the oscillators in either cell to be more aligned in phase. The smooth decrease of the phase lag with increasing coupling strength is shown in Fig. \ref{fig:variationwithcoupling}(b). In a similar way to the peak frequency, the phase lag between cells agrees with that predicted by LSA when the coupling is large.

Both of these factors, a shared oscillatory frequency and a minimal phase lag, are important for the proper functioning of a cellular clock. The changes in the peak frequencies and the phase lag that result from an increase in coupling strength correspond to quite a noticeable change in the quality of the oscillations themselves. Figs. \ref{fig:deterministicvsstochasticrh1point0} and \ref{fig:deterministicvsstochasticrh0point1} are evaluated for the same sets of parameters as Figs. \ref{fig:spectraphase}(a) and \ref{fig:spectraphase}(c) respectively. One parameter set is without coupling between the cells (Fig. \ref{fig:deterministicvsstochasticrh1point0}) and one is with cell-to-cell coupling (Fig. \ref{fig:deterministicvsstochasticrh0point1}). In the former case, the oscillations in either cell are somewhat aperiodic and there is no noticeable synchronisation between the cells. However, in Fig. \ref{fig:deterministicvsstochasticrh0point1}, the highs and lows of the cycles in either cell are more inclined to align -- this is associated with reduced phase lag and synchronised peak frequencies.
 
\subsubsection{Delta-Notch coupling increases the amplitude and coherence of noisy oscillations}\label{section:amplitudecoherence}
We asserted previously that an important characteristic of an effective cellular clock is a well-defined time-period -- if many frequencies contribute significantly to the oscillations, then it is more difficult to identify an overall phase for the clock. We also asserted the necessity of the cycles to have a significant amplitude. Both of these factors contribute to the clarity of the `signal' of the oscillations that constitute the cellular clock. We note that amplification and coherence are properties of the cycles in individual cells whereas synchronisation and phase lag are comparative measures of the oscillations in different cells. Despite this, amplification and coherence are indeed affected by Delta-Notch coupling too. 

\begin{widetext}

\begin{figure}[H]
\centering
        \includegraphics[scale = 0.18]{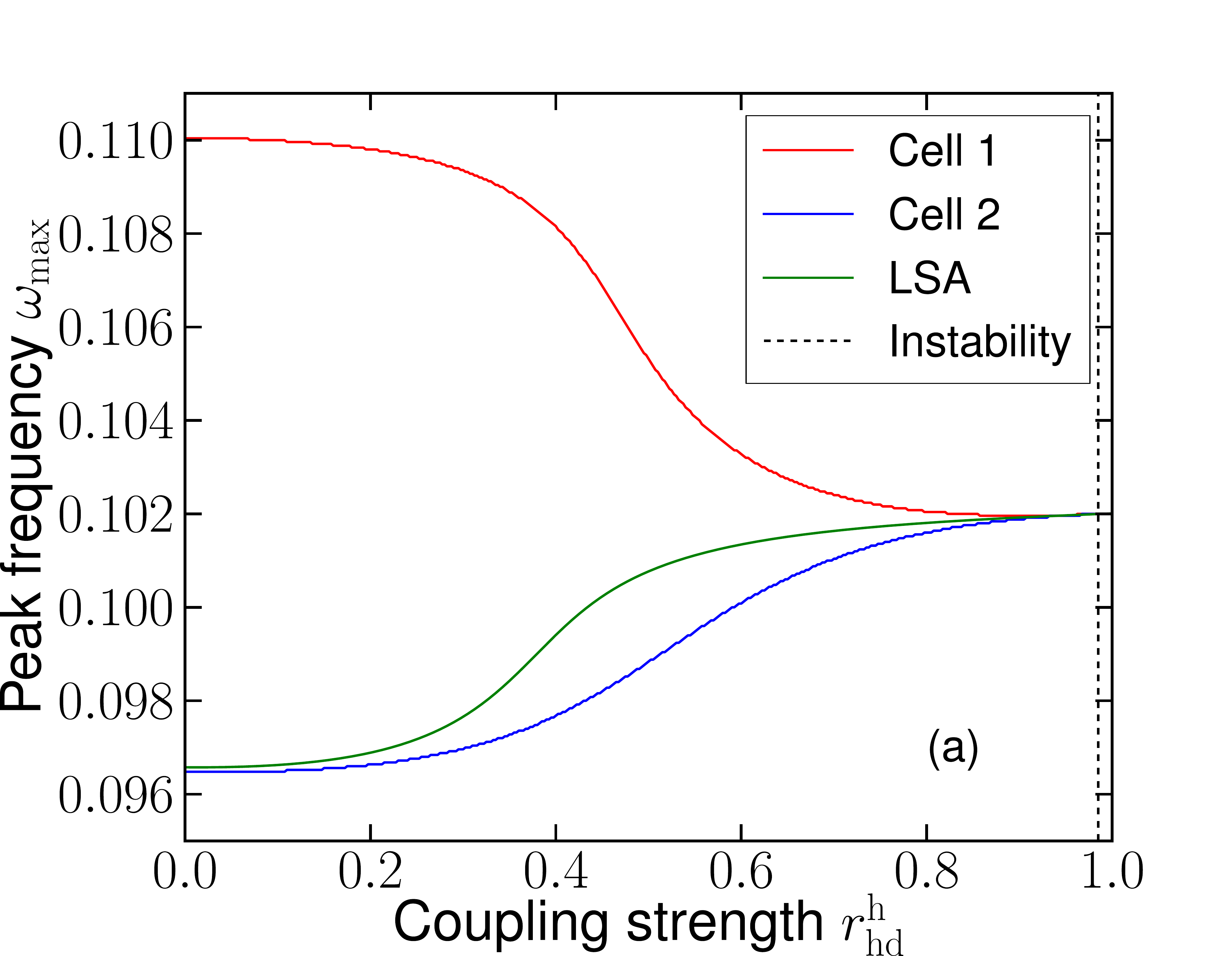}\hspace{2cm}
		\includegraphics[scale = 0.18]{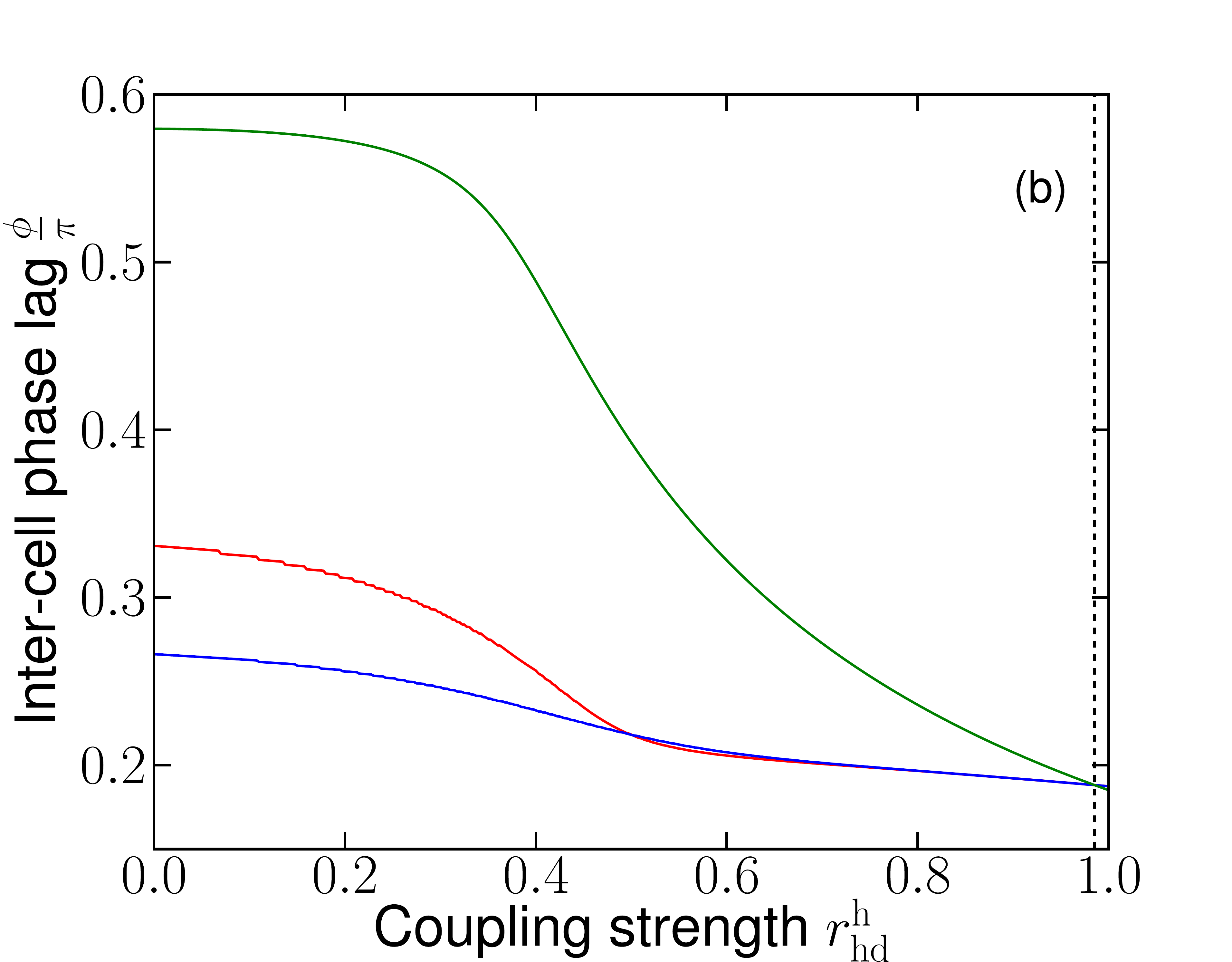}
		\\
        \includegraphics[scale = 0.18]{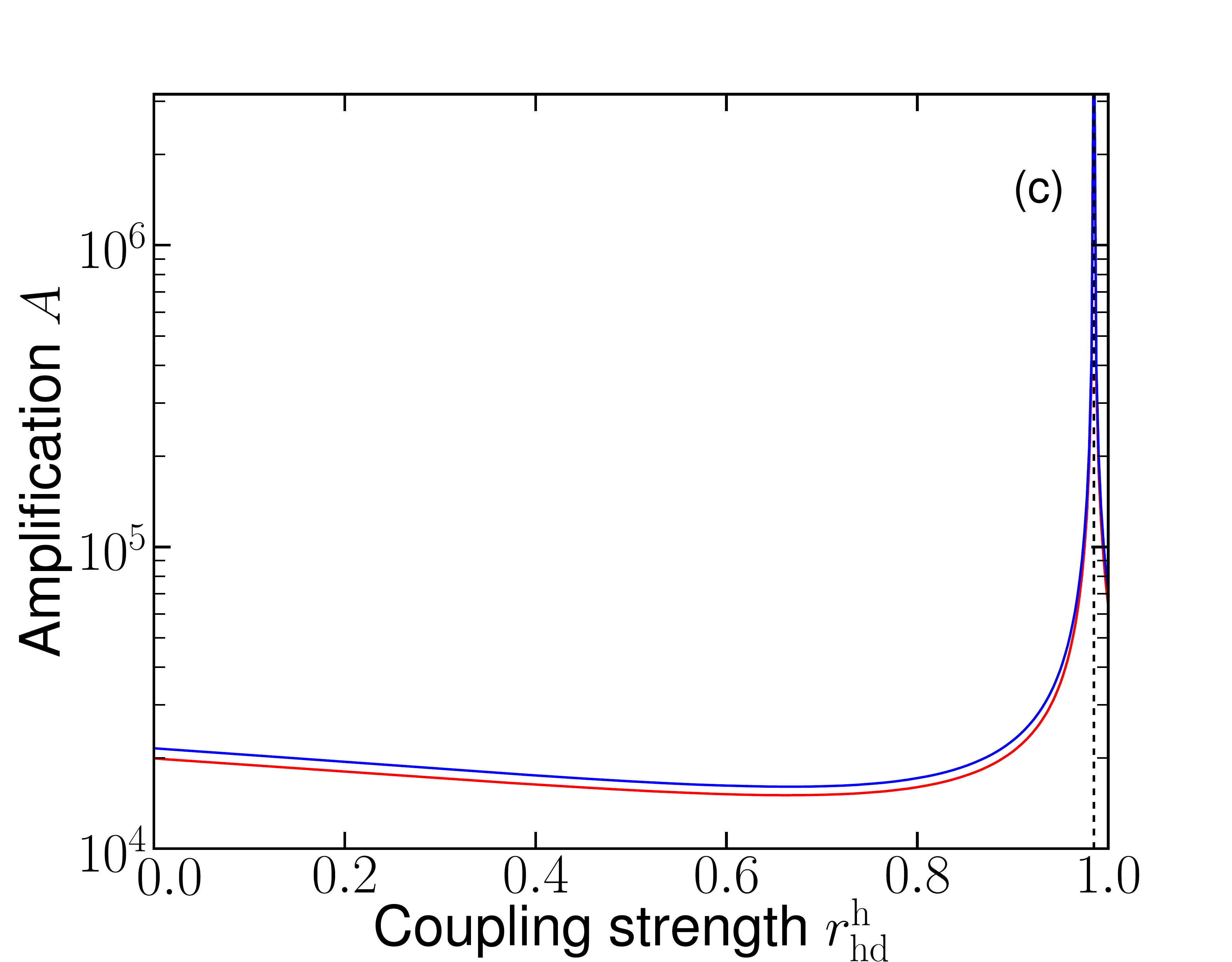}\hspace{2cm}
        \includegraphics[scale = 0.18]{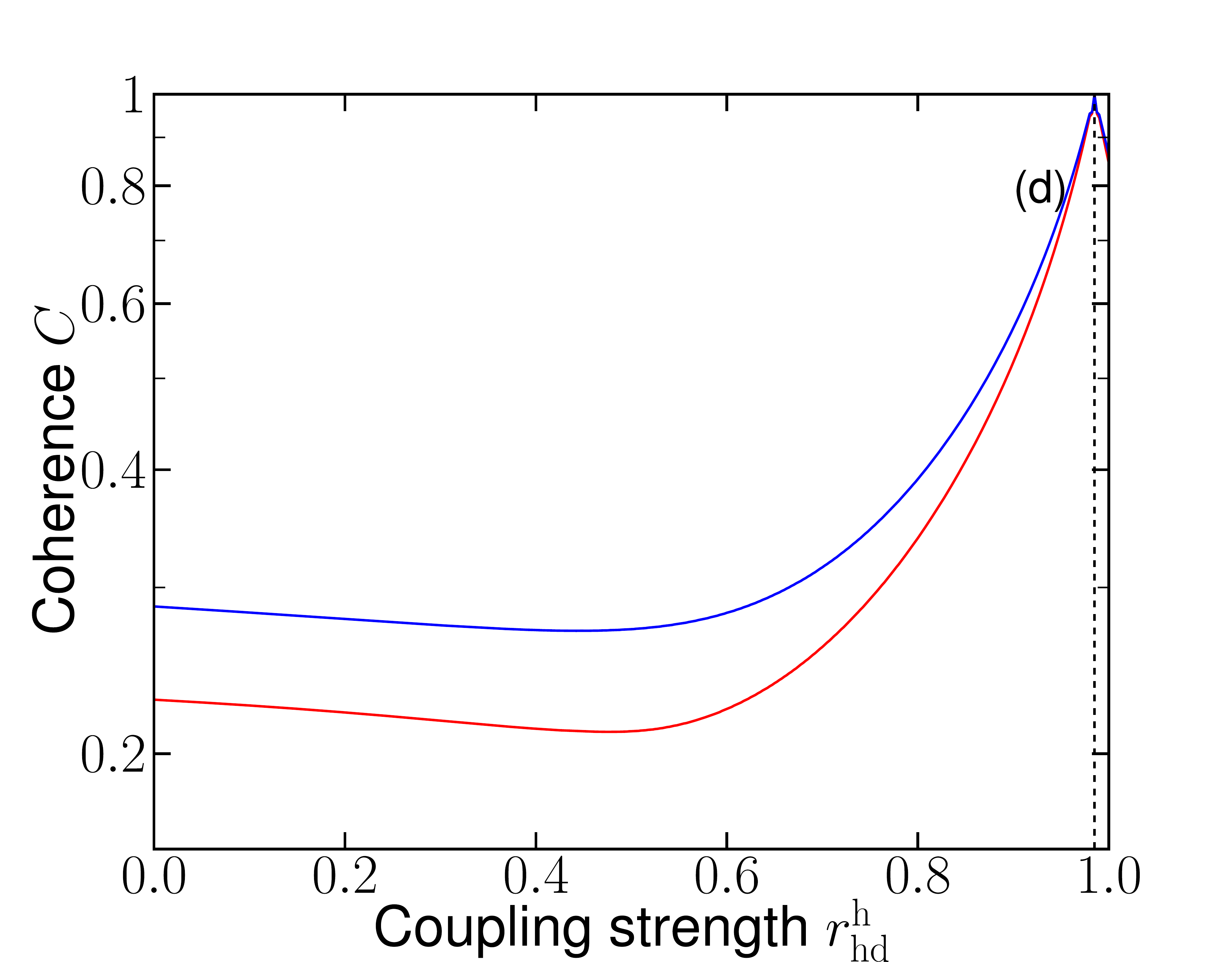}
        
	\caption{ Synchronisation (a), phase lag (b), amplification (c) and coherence (d) of oscillations of Hes/Her protein in the inhomogeneous two-cell system versus coupling strength produced using the linear theory detailed in Section S2 of the Supplemental Material. We use $\Delta\omega = 0.01$ in our definition of coherence (see Eq.~(\ref{coherencedef})). The system parameters are the same as in Fig. \ref{fig:deterministicvsstochasticrh1point0} but $r^{\mathrm{h}}_{\mathrm{hd}}$ and $r^{\mathrm{h}}_{\mathrm{h}}$ are varied subject to the constraint $r^{\mathrm{h}}_{\mathrm{h}} + r^{\mathrm{h}}_{\mathrm{hd}} = 1 $ (see Supplement Section S1 for the definition of these parameters). Panel (a) demonstrates that the peak frequency of oscillation $\omega^{(j)}_{\mathrm{max}}$ in either cell (where $j =1,2$ labels the two cells) approaches a common value as the coupling strength $r^{\mathrm{h}}_{\mathrm{hd}}$ is increased. This limiting value agrees with that predicted by linear stability analysis in the deterministic system (see Supplement Section S3). Panel (b) shows that as the two cells synchronise, the (rescaled) phase lag $\phi_{1,2}(\omega^{(j)}_{\mathrm{max}})/\pi$ between the two cells also decreases. Again, the value of the phase lag agrees with linear stability analysis for strong coupling. Panel (c) demonstrates how the oscillations in the two cells are initially dampened and subsequently amplified as the coupling strength is increased from zero. The theory becomes invalid as the deterministic fixed point of the system becomes unstable but is accurate close up to this point (see main text). Panel (d) shows how the coherence of the power spectra initially decreases and then increases as the coupling strength is increased, in a similar way to the amplification shown in panel (c). Values of the red and blue lines close to and to the right of the vertical dotted line at $r^{\mathrm{h}}_{\mathrm{hd}} = 0.985$ in all panels are not necessarily accurate; in this regime the system is close to or beyond the onset of the deterministic instability indicated by the vertical dashed line (emergence of a limit cycle). This results in corrections to the Fourier spectra which are not accounted for in our linear theory (see Supplement). It is at this point that the LSA, which is a deterministic analysis, becomes useful for identifying the peak frequency and phase lag. For low coupling strength (before the instability), the LSA is inaccurate because it does not take into account the effects of the noise, highlighting the need for the stochastic theory.  }
	\label{fig:variationwithcoupling}
\end{figure}

\end{widetext}

We find that as the Delta-Notch coupling strength is increased, the amplitude of the oscillations in both cells first decreases slightly then increases, as shown in Fig. \ref{fig:variationwithcoupling}(c). A similar result was also found in previous experimental and theoretical works \cite{ay, ozbudaklewis}. However, it is important to note the following caveat on the results presented in Fig. \ref{fig:variationwithcoupling}(c):

The blue and red lines shown in Fig. \ref{fig:variationwithcoupling} were produced using the theoretical approach based on the linear-noise approximation (see Sec. S2 of the Supplemental material). As such, we observe that the calculated amplitude of the oscillations diverges when the coupling strength is increased sufficiently. This singularity is a consequence of our theoretical approximation; it corresponds to the onset of an instability for the fixed point of the deterministic (noiseless) equations and the emergence of a limit cycle \cite{bolandgallamckane1, bolandgallamckane2}. The nonlinearities in the model then become more relevant and curtail the amplitude of the oscillations; this is not captured by the linear theoretical approach. The point at which this instability occurs is predicted by deterministic LSA (see Supplement Section S3) and is indicated by dashed vertical lines in Fig. \ref{fig:variationwithcoupling}. It is at this point that the stochastic theory, which is valid only when the deterministic system approaches a fixed point, becomes inaccurate. Instead, the LSA becomes the more accurate tool for identifying the peak frequency of oscillation and the inter-cell phase lag. The two analytical methods complement each other in this sense -- we can use both approaches to continuously analyse the dominant frequency and phase-lag over the onset of the deterministic instability. Unfortunately, the LSA provides no means of calculating the amplification or the coherence of the cycles -- we are only able to accurately predict these quantities when the deterministic dynamics approaches a fixed point (i.e. before the onset of the deterministic instability).  

With this caveat in mind, one can nevertheless see from Fig. \ref{fig:spectraphase} that the linear stochastic theory still agrees with simulation results close to this transition. That is, it remains accurate over a wide enough range of coupling strengths to faithfully capture an increase in the amplification with coupling strength, which occurs before the onset of the instability.

We find also that as the coupling strength is increased, the power spectrum of fluctuations becomes sharply peaked (as can be seen in Fig. \ref{fig:spectraphase}(c)) at a characteristic frequency, i.e., the cycles become more coherent (see Fig. \ref{fig:variationwithcoupling}(d)). The location of this peak in the power spectrum corresponds to the frequency to which the cycles in the two cells converge (as discussed in the previous section). 

We conceptualise this increase in amplification and coherence as a consequence of a kind of `resonant amplification'. Because of the communication between the cells, one can think of the state of one cell as influencing or `forcing' the oscillations in the neighbouring cells. As the inter-cell coupling strength is increased, the frequencies are aligned and the phase lag between them is reduced, the cycles begin to constructively interfere at a characteristic frequency.

Interestingly, as the coupling strength is increased from zero, the amplitude of the oscillations in either cell initially decreases (Fig. \ref{fig:variationwithcoupling}(c)), as does the peakedness of the power spectrum (Fig. \ref{fig:variationwithcoupling}(d)). We attribute this to the fact that, for low coupling, the oscillations in either cell are not adequately synchronised for their interference to be constructive. So as the coupling strength is increased initially, the `interference' between the two cells has a destructive effect. It is only when the phase lag is reduced and the frequencies are aligned sufficiently (as a result of a further increase of the coupling) that the collective amplitude of oscillations increases. 

The effects of increased amplification and coherence are evident in the time-series of the noisy cellular oscillations in the two-cell system shown in Figs. \ref{fig:deterministicvsstochasticrh1point0} and \ref{fig:deterministicvsstochasticrh0point1} (which are evaluated with and without inter-cellular coupling respectively). In Fig. \ref{fig:deterministicvsstochasticrh1point0}, there is no easily discernible periodic nature to the cycles in either cell. This is in contrast to the cycles in Fig. \ref{fig:deterministicvsstochasticrh0point1} where the highs and lows have more consistent temporal separations. This greater clarity in the oscillatory frequency is associated with the the increase in the sharpness of the peaks of the power spectra between Figs. \ref{fig:spectraphase}(a2) and \ref{fig:spectraphase}(c2) (i.e. an increase in coherence). Additionally, it can be seen that there are fewer pronounced highs and lows, on the whole, in the uncoupled system than in the coupled system; in Fig. \ref{fig:deterministicvsstochasticrh1point0} highs and lows are sporadically interrupted by stints of somewhat suppressed fluctuations about the fixed point. This in turn contributes to the lower overall amplification of the uncoupled system in comparison to that of the coupled system.

\subsection{Transcriptional/translational delays can lead to out-of-phase oscillations despite Delta-Notch coupling}\label{section:asynchronous}

We demonstrated in the previous sections that two cells with slightly disparate oscillatory frequencies could synchronise when coupled via Delta-Notch signalling. As the strength of the Delta-Notch coupling is increased, a common frequency is converged upon and the phase lag between the two cells is reduced. Although this is characteristic of many model parameter sets, it is not always the case. As was noted also in \cite{lewis} and \cite{momijimonk} (in the purely deterministic setting), neighbouring cells can be coupled in such a way that they oscillate in anti-phase with one another. This type of behaviour is facilitated by the delays associated with translation and/or transcription. 

Fig. \ref{fig:variationwithcouplinganti} demonstrates that for certain parameter sets, as the coupling is increased, oscillations in two cells can converge on a common frequency but the phase lag between the two cells can approach values closer to $\phi = \pi$ than $\phi = 0$. That is, the cells tend towards oscillating in anti-phase with one another. Clearly, this is suboptimal if these cellular oscillations are to be used as a segmentation clock.

\begin{widetext}

\begin{figure}[H]
\centering
        \includegraphics[scale = 0.18]{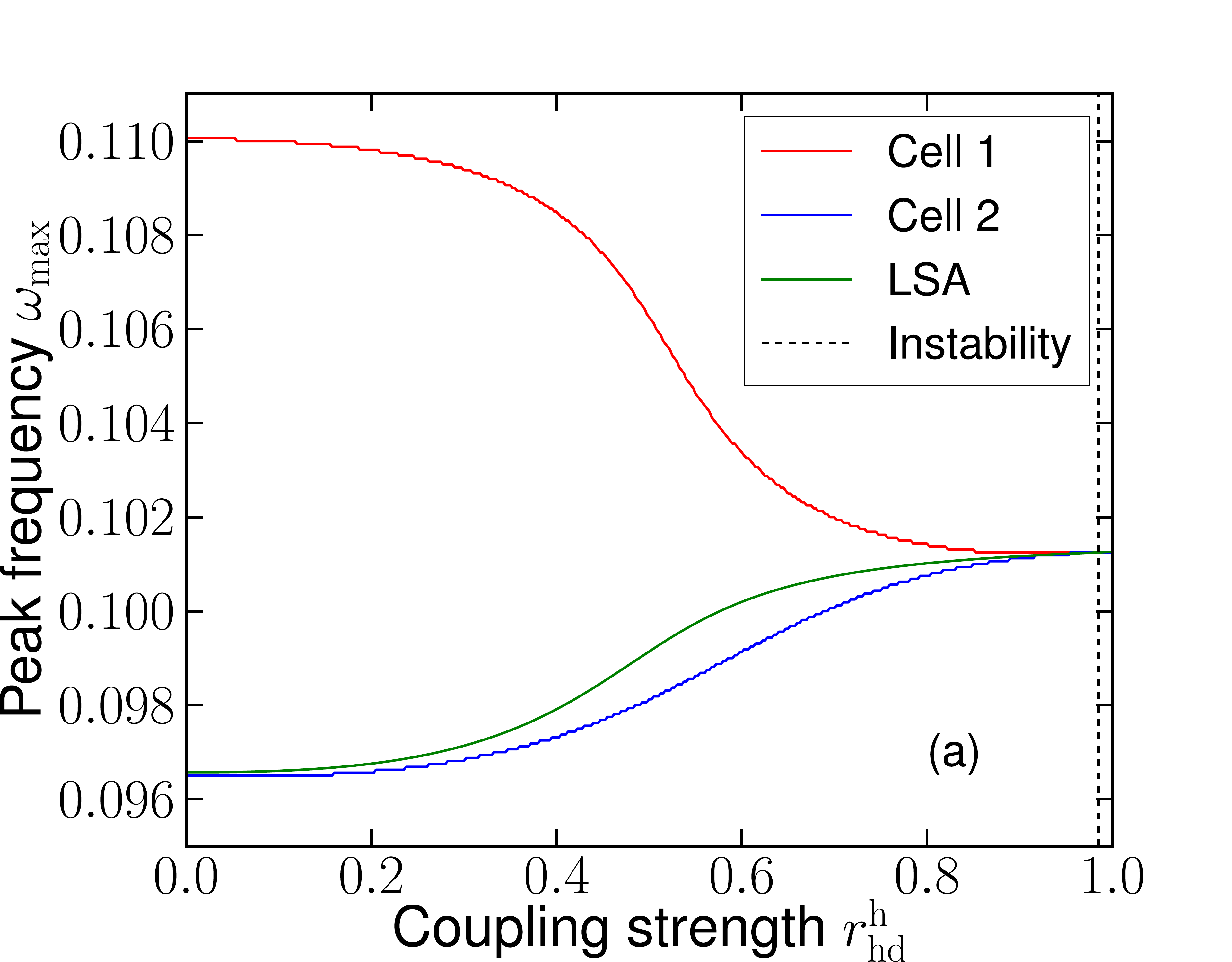}\hspace{2cm}
		\includegraphics[scale = 0.18]{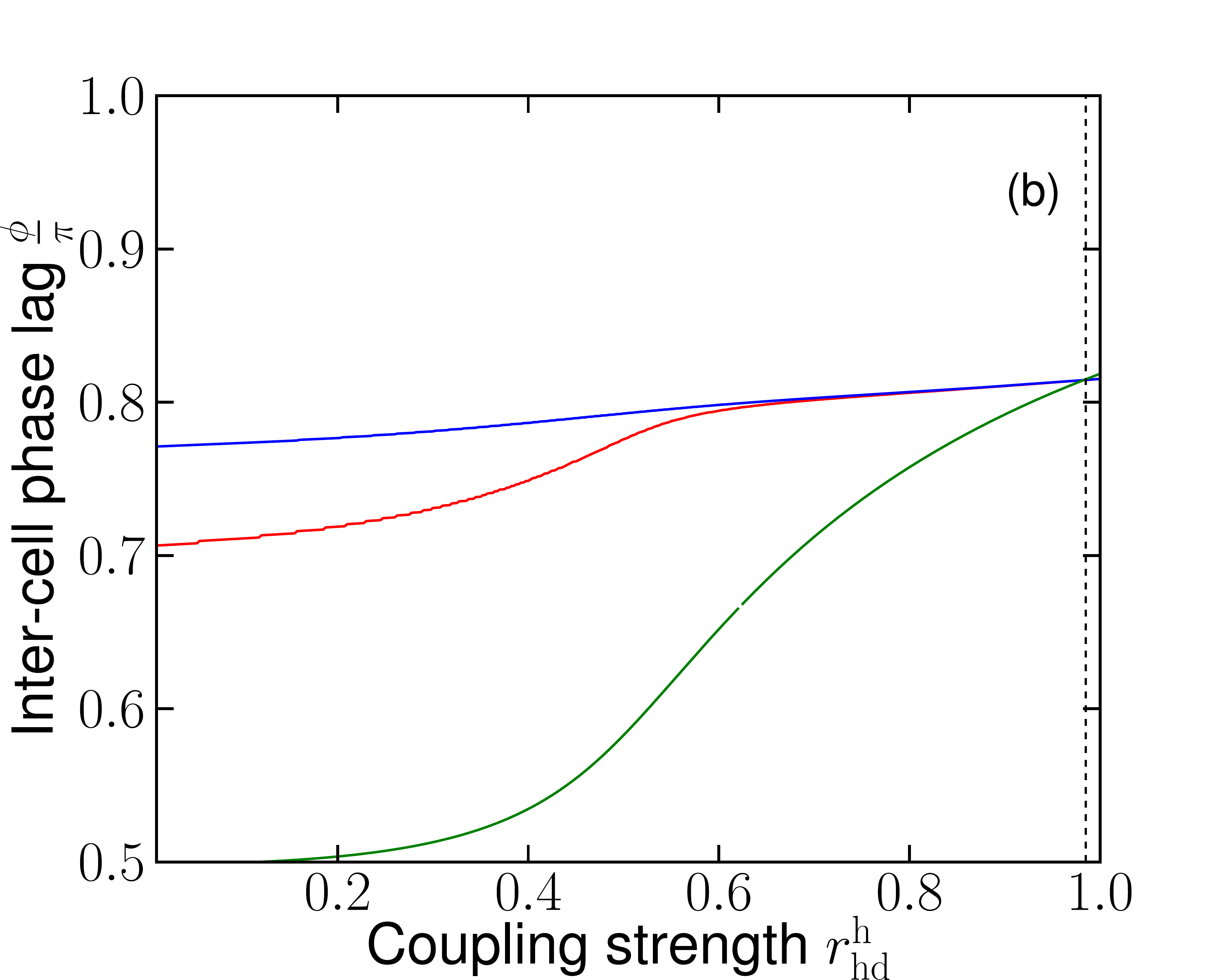}
        
	\caption{Peak frequencies $\omega^{(j)}_{\mathrm{max}}$ and rescaled phase lag $\phi_{1,2}(\omega^{(j)}_{\mathrm{max}})/\pi$ in the two-cell system (where $j =1,2$ labels the two cells) as a function of the coupling strength for a parameter set which results in anti-phase synchronisation. The system parameters are the same as in Fig. \ref{fig:deterministicvsstochasticrh1point0} but $r^{\mathrm{h}}_{\mathrm{hd}}$ and $r^{\mathrm{h}}_{\mathrm{h}}$ are varied subject to the constraint $r^{\mathrm{h}}_{\mathrm{h}} + r^{\mathrm{h}}_{\mathrm{hd}} = 1 $ and $\tau^{(m)d} = 20$. Panel (a) demonstrates that as the coupling strength $r^{\mathrm{h}}_{\mathrm{hd}}$ is increased, the peak frequency of oscillations approaches a common value, as in Fig. \ref{fig:variationwithcoupling}. However, panel (b) shows that as the two cells synchronise, the phase lag between the two cells increases, tending towards $\pi$ instead of $0$. }
	\label{fig:variationwithcouplinganti}
\end{figure}

To understand why this should happen, we monitor the dominant frequency and the associated phase difference between the two cells as the transcriptional/translational delays are varied (see Fig. \ref{fig:synchswitching}). For the purposes of this analysis, the two cells are taken to be identical. In this case, the two cells are guaranteed to share a peak oscillatory frequency $\omega_{\mathrm{max}}^{(1)}=\omega_{\mathrm{max}}^{(2)}=\omega_{\mathrm{max}}$. We find that whether the cells oscillate in or out-of phase is determined by an interplay between the delays and the dominant frequency of oscillation.

\begin{figure}[H]
	\centering
        \includegraphics[scale = 0.2]{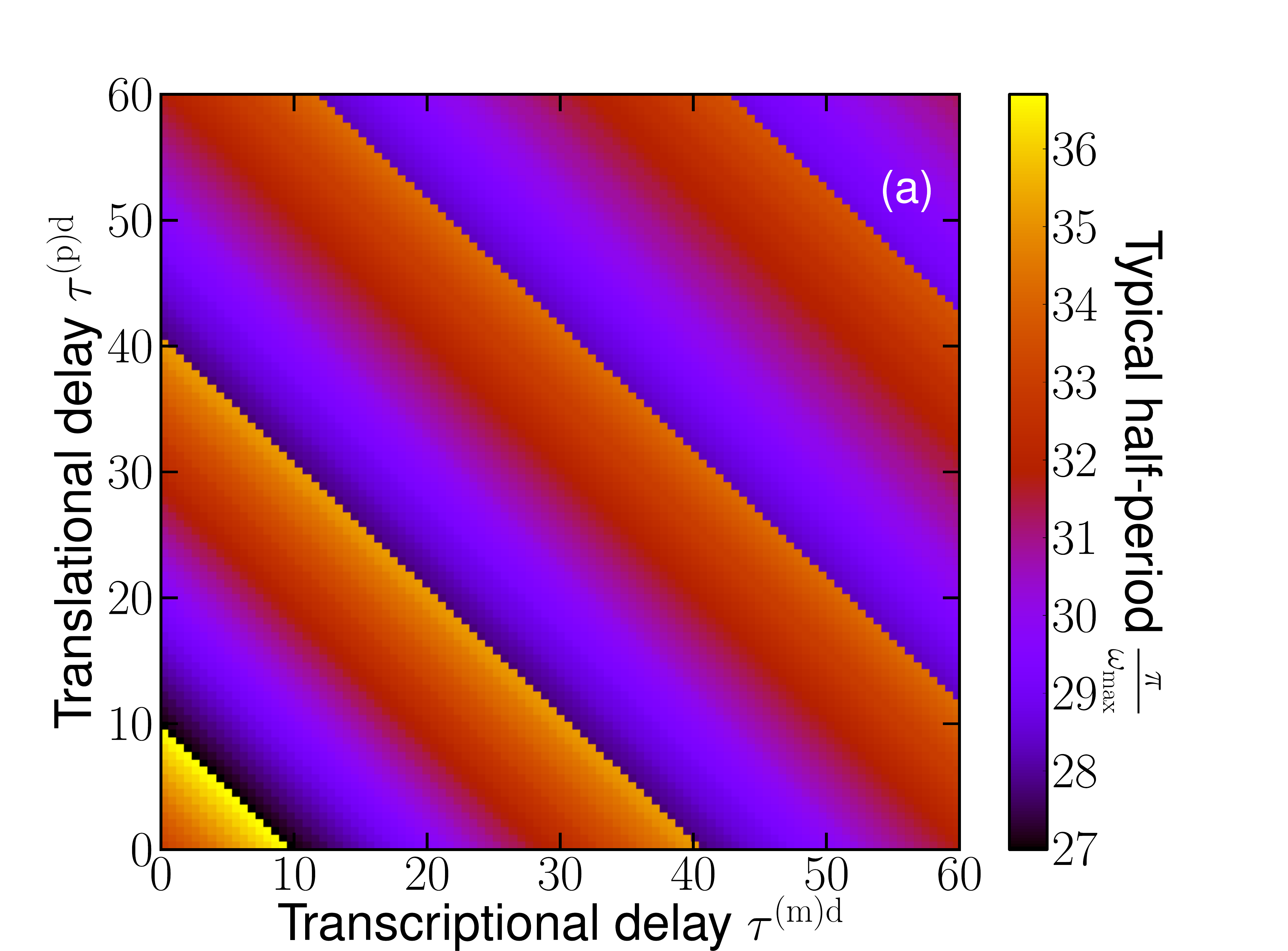}
		\hfill
		\includegraphics[scale = 0.2]{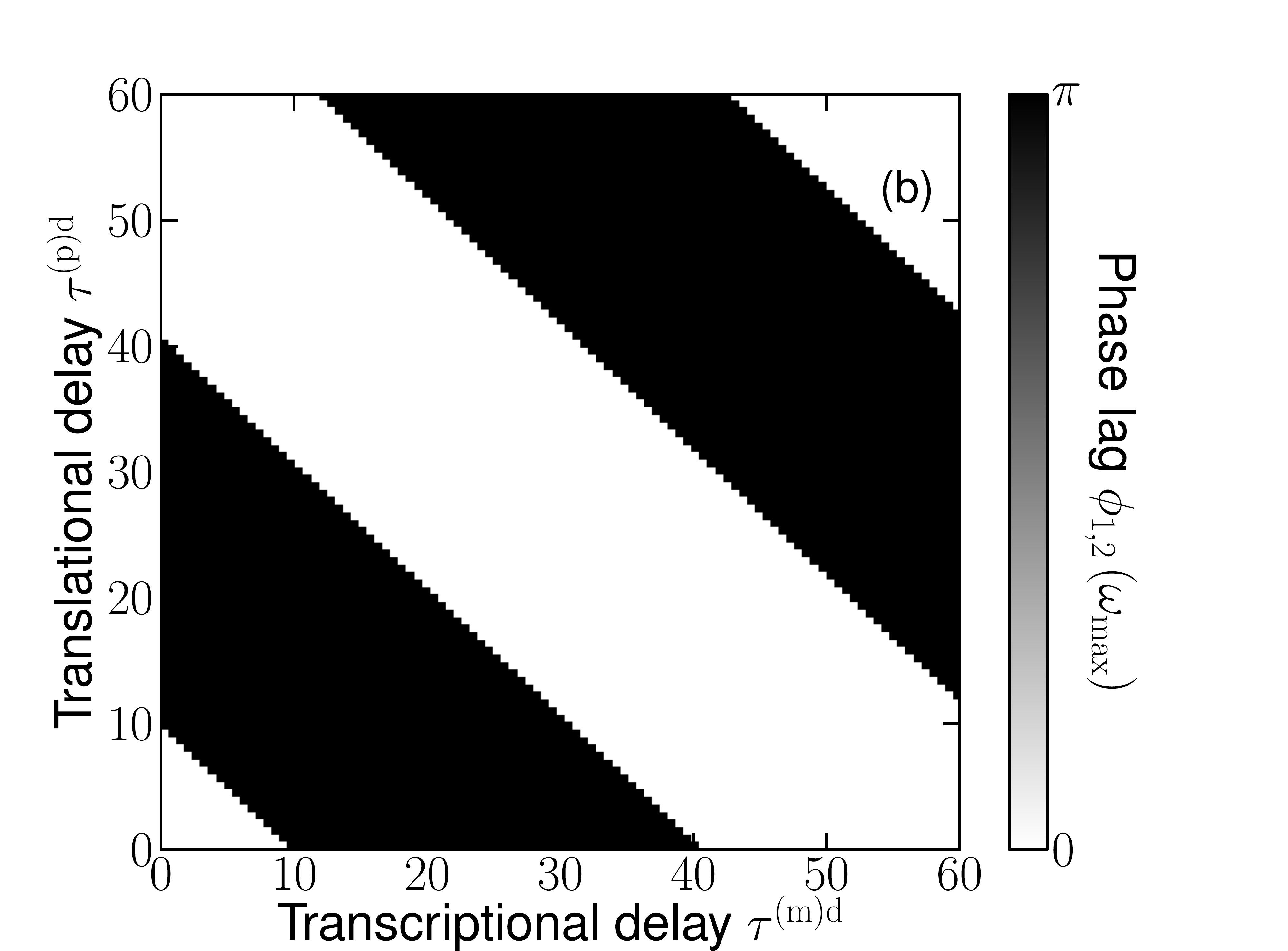}
	\caption{Half the time period associated with dominant frequency of oscillation in the 2-cell system $T/2= \pi/\omega_{\mathrm{max}}$ (a) and the corresponding phase lag between the cells (b). The system parameters are the same as in Fig. \ref{fig:deterministicvsstochasticrh0point1} but $\tau^{(m)h} = 20$ in both cells and $\tau^{(p)d}$ and $\tau^{(m)d}$ are varied. The typical half-period varies around an average value of $\sim 30 \mathrm{mins}$. Although $T/2$ is dependent on the time delays $\tau^{(p)d}$ and $\tau^{(m)d}$, it typically remains within $\sim 10 \%$ of this mean value. The phase lag switches between $0$ and $\pi$ along the lines lines $\tau^{(p)d}+\tau^{(m)d} = \tau^{\mathrm{tot}}_i$ (where $i = 1,2,3,\dots$). The values $\tau^{\mathrm{tot}}_i$ at which the switch in phase occurs are separated by regular intervals such that $\tau^{\mathrm{tot}}_{i}-\tau^{\mathrm{tot}}_{i-1} \approx 30 \mathrm{mins}$ -- roughly the typical half-period.}
	\label{fig:synchswitching}
\end{figure}

\end{widetext}

One observes that the phase lag between the two cells switches between $0$ and $\pi$ (and vice versa) when the total \textit{delta} delay time $\tau^{\mathrm{tot}} = \tau^{(m)d}+\tau^{(p)d}$ reaches certain values $\tau^{\mathrm{tot}}_i$, where $i=1,2,3,\dots$ (we notice a similar effect when other pairs of delays are varied). In Figure \ref{fig:synchswitching}, the times at which the switches occur are separated by a regular time interval such that $\tau^{\mathrm{tot}}_i-\tau^{\mathrm{tot}}_{i-1} \approx 30 \mathrm{mins}$.  This interval is roughly equal to half of the typical period of the oscillations, $T/2 = \pi/\omega_{\mathrm{max}}$, which varies within around $\sim 10\%$ of the mean value $\sim 30 \mathrm{mins}$. 

In order to gain some intuition for what this means, we suppose that for a particular set of delays, the two cells oscillate in phase. If we add an additional $\pi/\omega_{\mathrm{max}}$ to the total delay time, the `signal' from one cell to its neighbour is delayed by half a cycle. If the cells would oscillate in phase without this additional delay, it stands to reason that they would oscillate in anti-phase given the additional delay -- there would be no difference from the point of view of either cell (assuming that the frequency of oscillation does not vary greatly with the changing delay times). A similar argument holds true for the addition of $2\pi/\omega_{\mathrm{max}}$ to $\tau^{\mathrm{tot}}$ -- in this case the phase difference between the two cells ought to be unchanged. One caveat to this reasoning is that the frequency of the oscillations cannot be changed too drastically by the variation in time delay. 

We conclude that while the time delays associated with translation and transcription are crucial for the persistence of the cycles which constitute the cell clock, they are somewhat of a double-edged sword. Depending on the interaction between the delays and the internal clocks of each cell, the cells may oscillate in anti-phase with one another. As a result, the precise nature of the transcriptional/translational delays is of great importance with regards to the proper synchronisation of the segmentation clock.

\subsection{Transcriptional/translational delays can disrupt global oscillations in chains of cells and give rise to waves of gene expression}\label{section:waves}

In this section, we discuss how the preceding analysis can be extended to a chain of Delta-Notch-coupled cells. We demonstrate that synchronised noisy oscillations in the two-cell system can correspond to global oscillations in a chain. We also explore behaviours other than global oscillations which can occur as a result of delays; namely, the emergence of noise-induced waves. 

A distinction between oscillations of the deterministic trajectory and purely noise-induced oscillations was made in Section \ref{section:deterministicmodels}. In a similar way, one finds that waves can manifest in the individual-based system when they do not in the deterministic system. Such `stochastic waves' or `quasi-waves' have been found previously in theoretical models of individual-based systems with long-range interaction \cite{biancalanigalla}. Here however, the stochastic waves arise due to a combination of the non-local dependence of the reaction rates (due to Delta-Notch signalling) and the transcriptional/translational delays. Conversely, waves of gene expression have been studied previously in chains of coupled genetic oscillators \cite{momijimonk} but this was done in the context of deterministic equations which ignored intrinsic noise.

We mention the emergence of waves here not so much as an explanation for the travelling waves which are seen in the PSM (these are most likely due to a variation in translatonal/transcriptional delay along the anterior-posterior axis \cite{keskin, ay}), but as an illustration of the different kinds of undesirable behaviour which can arise when cells oscillate out of phase with one another. As such, the results presented in section are exploratory and not necessarily an attempt to recreate any (as yet) observed phenomena.

We find that for sets of parameters where one would observe oscillations in anti-phase in the two-cell system, one finds waves of gene expression in an extended chain of cells. For parameter sets where the cycles in the two-cell system oscillate in unison, one observes global in-phase oscillations in gene expression.

For an extended chain of cells, we define the power spectrum 
\begin{align}
P_k^\alpha\left(\omega\right) = \langle \lvert\hat{\tilde\delta}^\alpha_k\left(\omega\right) \rvert^2 \rangle ,  \label{powerspectraspatial}
\end{align}
where we have used the discrete Fourier transform with respect to the cell number $j$ defined by $\tilde f_{k} = \frac{1}{\sqrt{L}}\sum_j e^{ij k} f_j $, where $L$ is the number of cells in the chain. Details of the calculation of this power spectrum are given in the Supplement (Section S2 B).

In a chain of coupled cells, global in-phase oscillations are characterised by a peak in the power spectrum $P_k^\alpha\left(\omega\right)$ at spatial wavenumber $k=0$ and non-zero oscillatory frequency $\omega$. Such a power spectrum is shown in Fig. \ref{fig:synchedcycles}(a), and an example of the corresponding behaviour in a chain of cells is demonstrated in Fig. \ref{fig:synchedcycles}(b), where one observes that the peaks and troughs in one cell tend to align with those in the neighbouring cells. That the cells are indeed in phase with one another is verified by the phase lag (see inset of Fig. \ref{fig:synchedcycles}(a)) which is equal to zero, regardless of cell separation. 

Stochastic waves, on the other hand, are characterised by a peak in the Fourier power spectrum at non-zero values of both the spatial wavenumber $k$ and the temporal frequency $\omega$. An example of such a power spectrum is given in Fig. \ref{fig:travellingstochasticwaves}(a). To validate the claim that this peak in the power spectrum is indicative of travelling waves, we note that the phase difference between cells varies linearly with cell separation, as is shown in the inset of Fig. \ref{fig:travellingstochasticwaves}(a). We stress that the coupling between the cells here is biased in one direction, which breaks the symmetry of the system, allowing waves to travel. Travelling waves of gene expression have been observed in experimental systems other than the PSM \cite{wendenleaf, kazuyalettuce}.

For symmetric coupling however, one instead observes standing waves of gene expression, where alternate cells oscillate in antiphase, as shown in Fig. \ref{fig:standingstochasticwaves}. In this particular case, the phase lag between any pair of adjacent cells (at the peak frequency) is $\pi$, as can be seen in the inset of Fig. \ref{fig:standingstochasticwaves}(a). This is rather reminiscent of the on-off chequerboard patterns associated with neural differentiation and lateral inhibition \cite{mainilateralinhibition}. 

Examples of travelling and standing stochastic waves in a chain of coupled cells are shown in Figs. \ref{fig:travellingstochasticwaves}(b) and \ref{fig:standingstochasticwaves}(b) respectively. There is a clear qualitative distinction between the two. In Fig. \ref{fig:travellingstochasticwaves}(b), peaks and troughs in one cell gradually travel in the positive $j$ direction as time goes by, indicating a travelling wave. In Fig. \ref{fig:standingstochasticwaves}(b) however, the peaks in one cell tend to line up with the troughs of the neighbouring cells and visa versa and there is no clear direction of travel. 

\begin{widetext}

\begin{figure}[H]
	\centering
        \includegraphics[scale = 0.207]{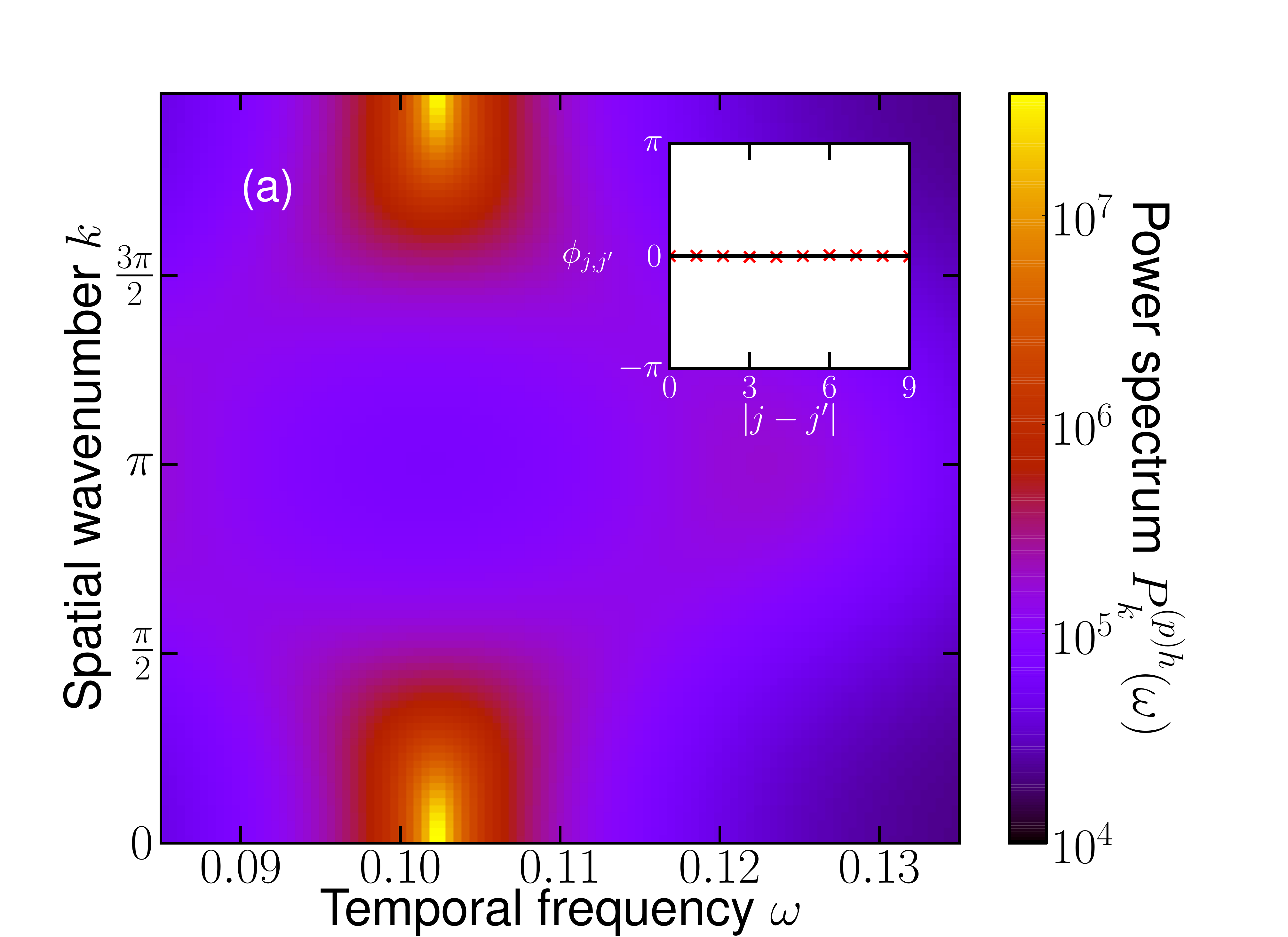}
		\hfill
		\includegraphics[scale = 0.209]{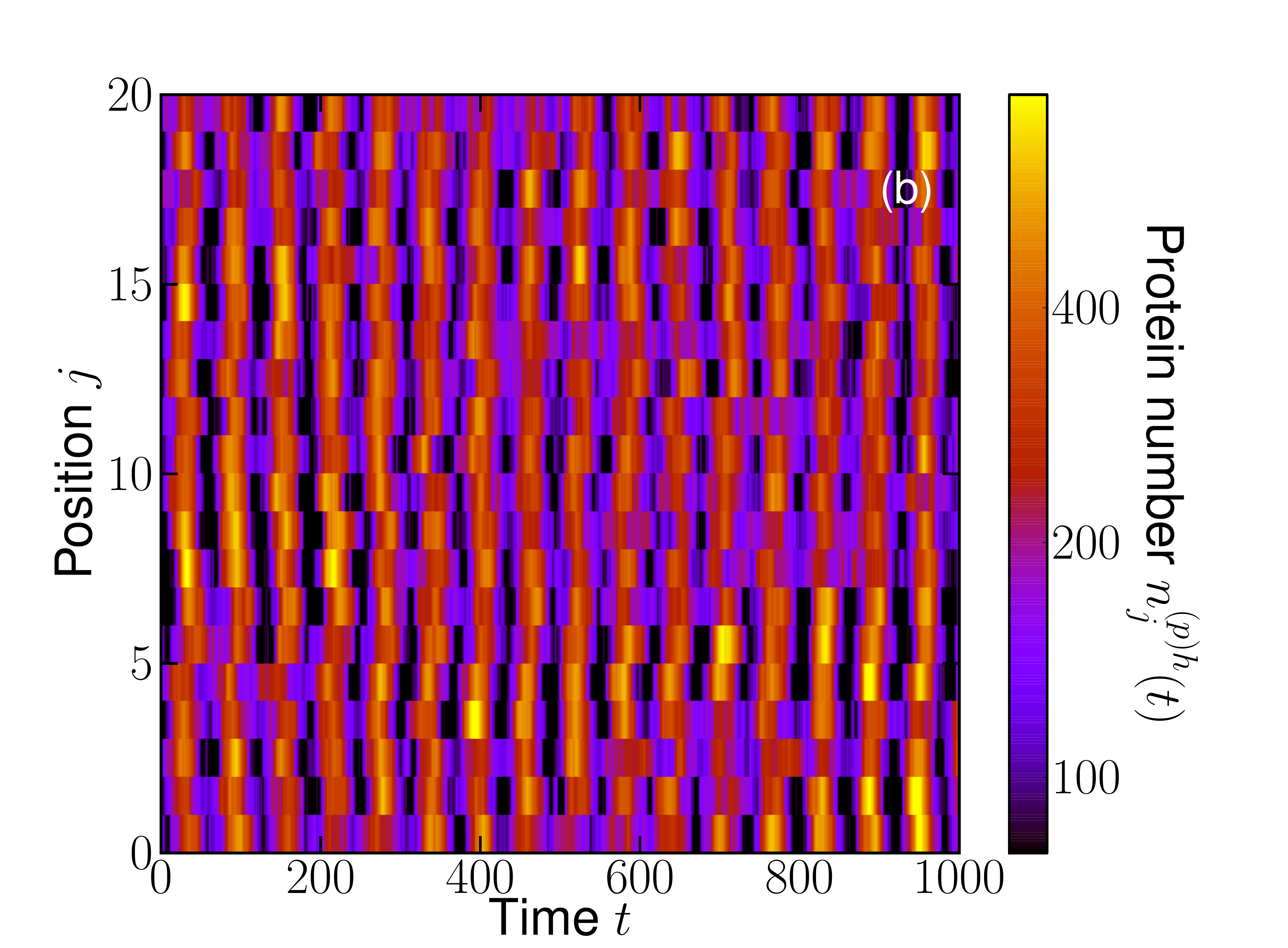}
	\caption{ Global oscillations in Hes/Her protein numbers. Panel (a): The power spectrum of fluctuations $P^{(p)h}_k\left(\omega\right)$ and the corresponding phase lag between cells at peak frequency  $\phi^{(p)h}_{j,j'}\left(\omega_{\mathrm{max}}\right)$ as a function of separation (inset). Global oscillations are indicated by the fact that the peak of the power spectrum is located at $k=0$ and non-zero $\omega$ and also by the phase lag between cells being zero. Panel (b): An example of noise-induced global oscillations. Data is from a simulation of the stochastic individual-based model. The peaks and troughs in \textit{hes/her} expression have a tendency to align, giving rise to vertical striped structures in the figure. The coupling between cells is symmetric ($d^{(+)} = d^{(-)}= 1$). The remaining system parameters are the same as in Fig. \ref{fig:deterministicvsstochasticrh0point1} with the exception that $\tau^{(m)h} = 20$ in all cells. }
	\label{fig:synchedcycles}
\end{figure}

\begin{figure}[H]
	\centering
        \includegraphics[scale = 0.209]{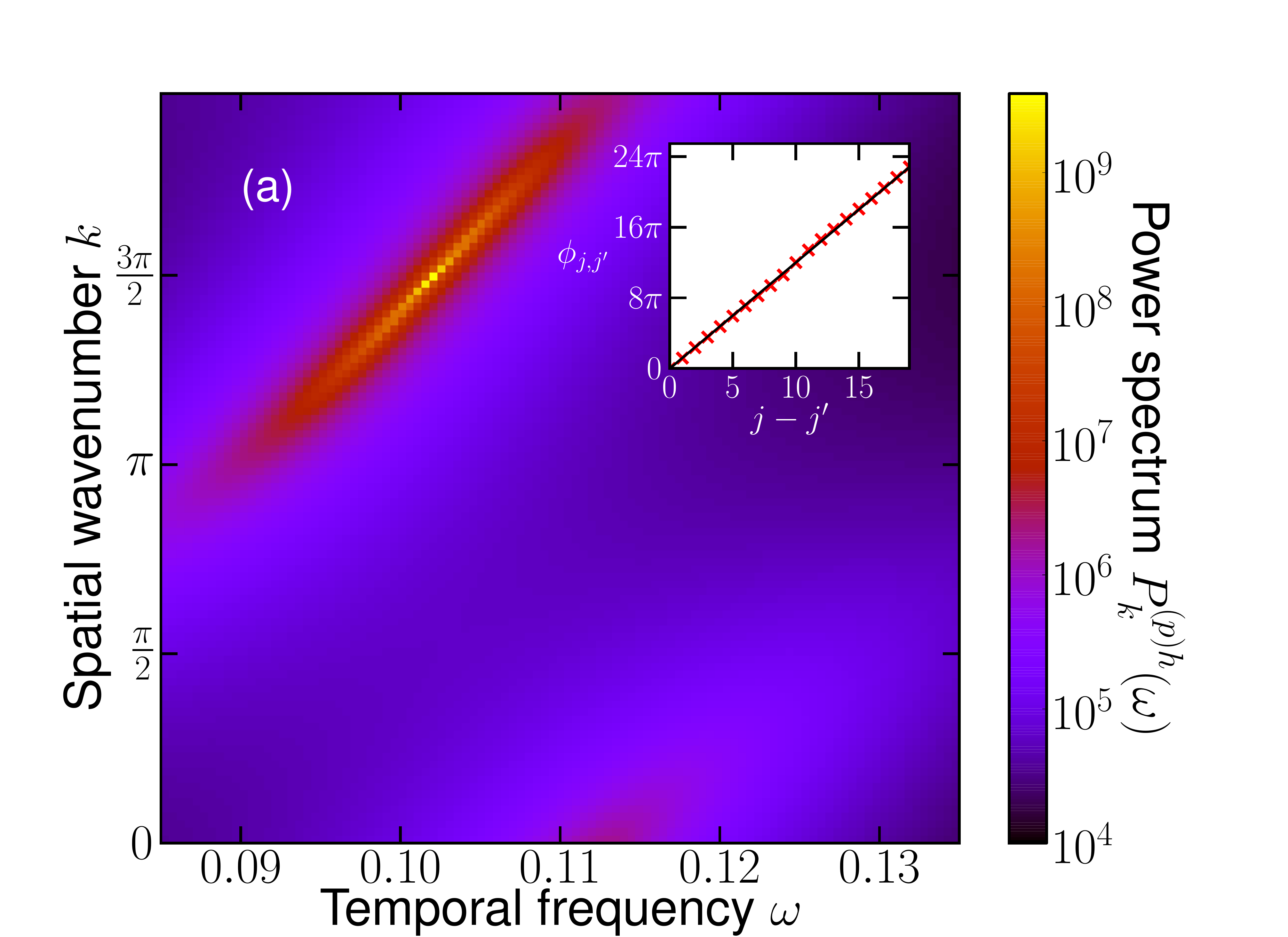}
		\hfill
		\includegraphics[scale = 0.207]{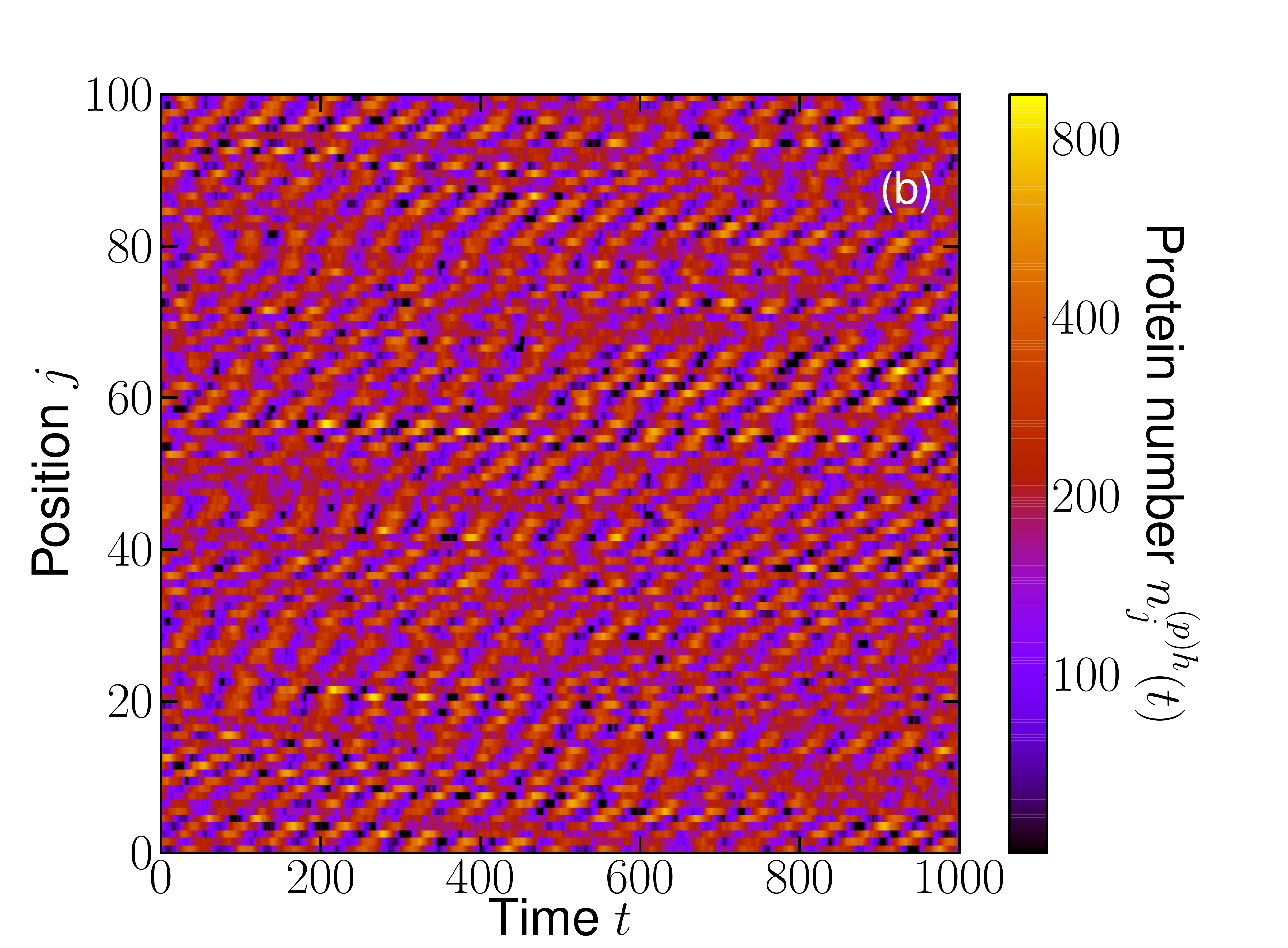}
	\caption{Travelling stochastic waves. Panel (a): The power spectrum of fluctuations $P^{(p)h}_k\left(\omega\right)$ and the corresponding phase lag between cells at peak frequency  $\phi^{(p)h}_{j,j'}\left(\omega_{\mathrm{max}}\right)$ as a function of cell separation (inset). The presence of travelling waves is indicated by a sharp peak in the power spectrum at a non-zero value of $\omega$ and a value of $k$ that is neither equal to $0$ nor $\pi$. The presence of travelling waves is further evidenced by a phase lag which is linearly increasing with cell separation. We note that we have relaxed the condition that $\phi^{(p)h}_{j,j'}$ be in the range $[-\pi, \pi)$ in order for this trend to be apparent. Panel (b): An example of noise-induced travelling waves; this is from simulations of the stochastic individual-based model. When a peak or a trough occurs in one cell, it has a tendency to move upwards to the neighbouring cell as time progresses, giving rise to the diagonal structures in the figure. The coupling between cells is asymmetric ($d^{(+)} = 2$, $d^{(-)}= 0$), which breaks the directional symmetry of the system, allowing waves to propagate. The remaining system parameters are the same as in Fig. \ref{fig:deterministicvsstochasticrh1point0} with the exception that $r^h_h = 0.07$, $r_{hd}^h = 0.93$, $\tau^{(m)h} = 20$ in all cells and $\tau^{(m)d} = 35$.}
	\label{fig:travellingstochasticwaves}
\end{figure}

\begin{figure}[H]
	\centering
        \includegraphics[scale = 0.207]{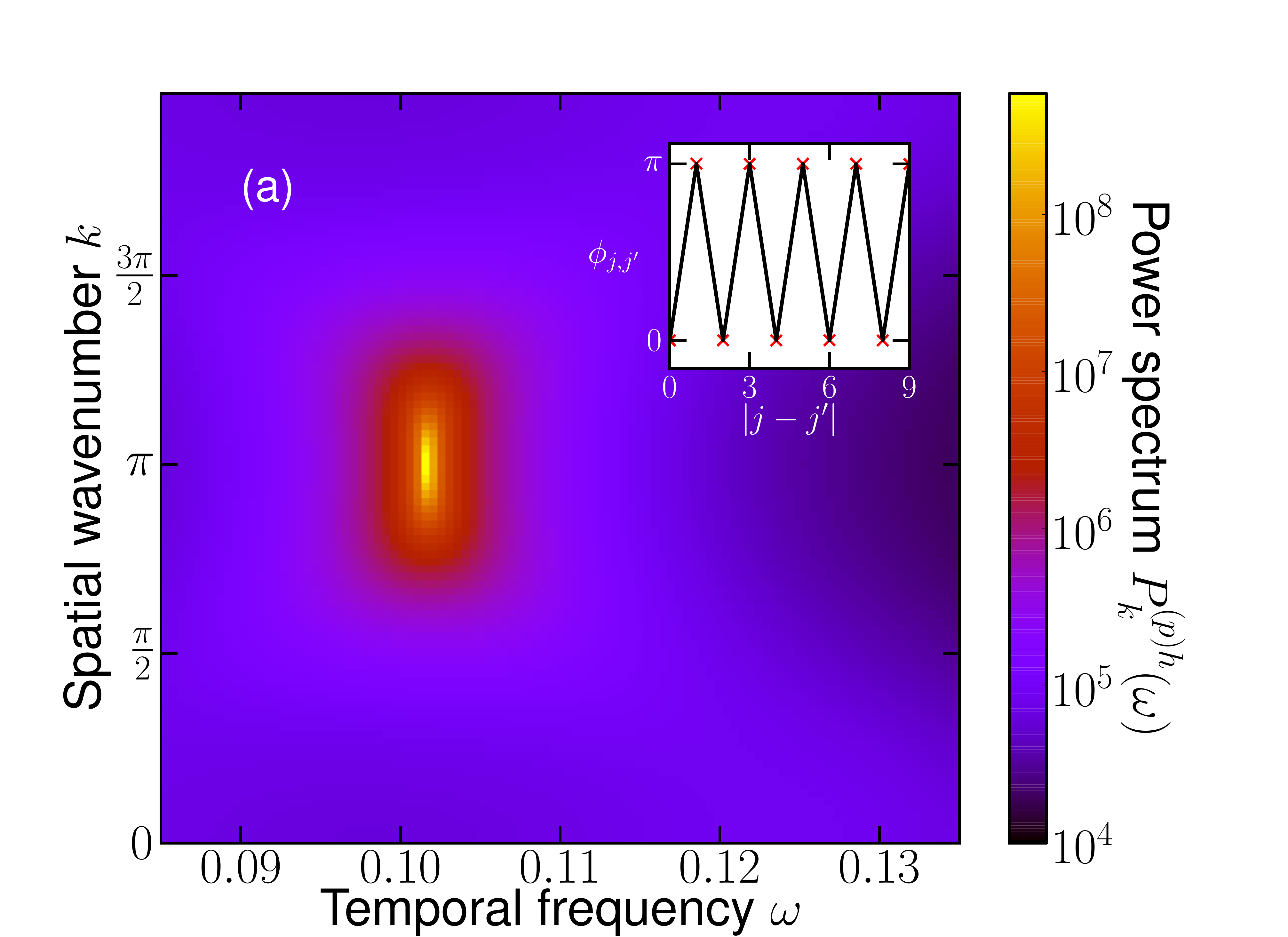}
		\hfill
		\includegraphics[scale = 0.209]{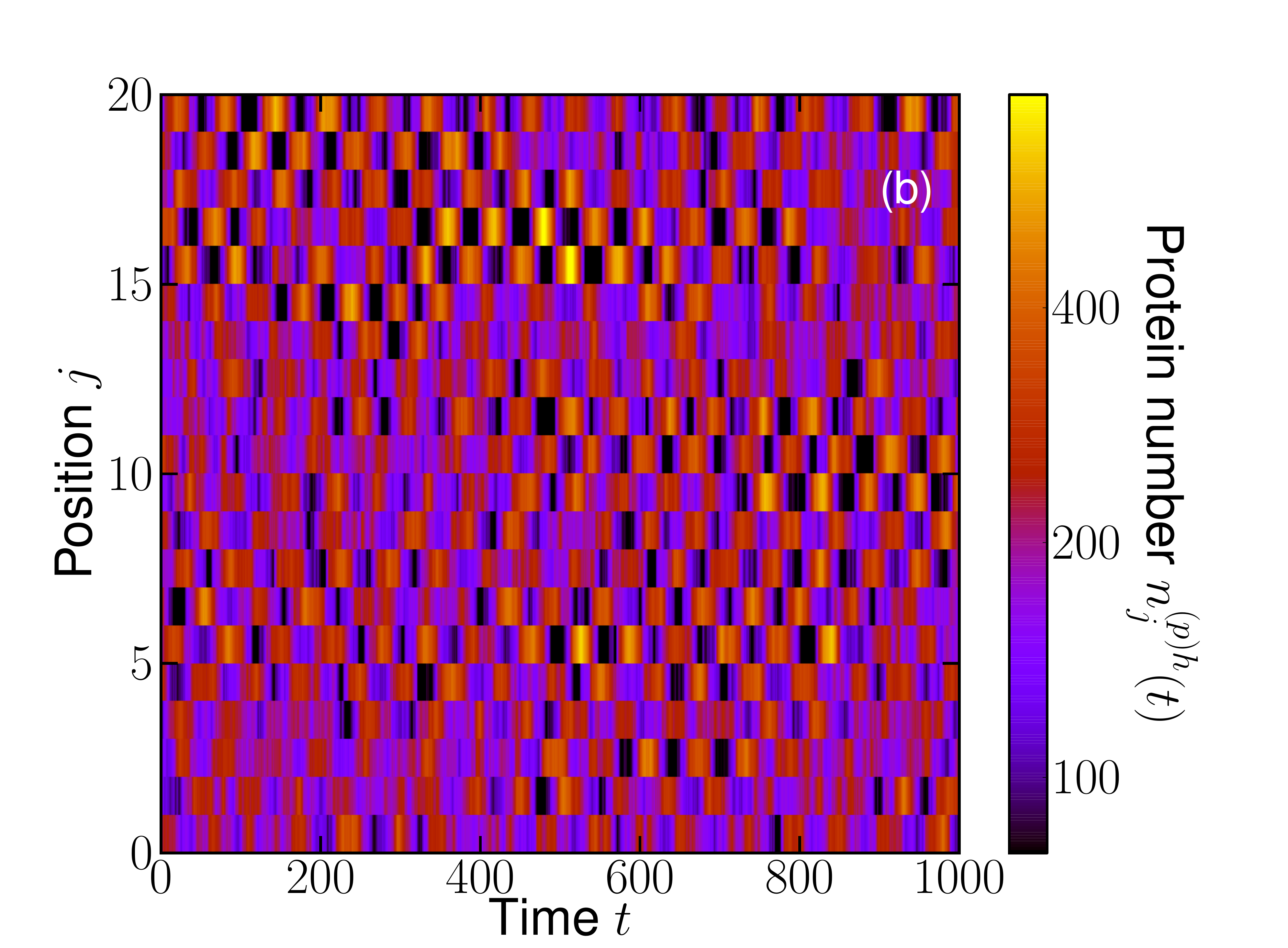}
	\caption{Neighbouring cells oscillating in anti-phase (standing waves). Panel (a): The power spectrum of fluctuations $P^{(p)h}_k\left(\omega\right)$ and the corresponding phase lag between cells at peak frequency  $\phi^{(p)h}_{j,j'}\left(\omega_{\mathrm{max}}\right)$ as a function of separation (inset). Neighbouring cells tend to oscillate in anti-phase with one another, as demonstrated by the phase lag (inset) and also by the location of the peak of the power spectrum at $k= \pi$ and non-zero $\omega$. Panel (b): An example of noise-induced standing waves seen in simulations of the stochastic individual-based model. The peaks in one cell tend to align with the troughs in the neighbouring cell,  giving rise to chequerboard-type patterns in the figure. The coupling between cells is symmetric ($d^{(+)} = d^{(-)}= 1$). The remaining system parameters are the same as in Fig. \ref{fig:deterministicvsstochasticrh1point0} with the exception that $r^h_h = 0.07$, $r_{hd}^h = 0.93$, $\tau^{(m)h} = 20$ in all cells and $\tau^{(m)d} = 20$. }
	\label{fig:standingstochasticwaves}
\end{figure}

\end{widetext}

\section{Summary and discussion}\label{section:conclusion}
The process of somite segmentation poses a complex many-faceted problem for theorists and experimentalists alike and remains an area of active inquiry. Much is left to be discovered about the precise nature of the role of each of the genes involved with the somite segmentation clock and their interactions with external signalling factors in the embryo.

In this paper, we aim to have provided some insight into the role that Delta-Notch signalling plays in not only aligning the frequencies of oscillation of cyclic gene expression in neighbouring cells but also in reducing phase lag, in improving the coherence of oscillations and in increasing their amplitude, so as to produce a robust and reliable segmentation clock. We also explored the role that intrinsic noise plays in the system; counter to intuition, it can actually promote persistent cycles, rather than obscure them. Further, we discussed how the delays involved in the transcription and translation processes can act to promote oscillations but can also result in neighbouring cells oscillating out-of-phase with one another. We examined how this resulted from an interplay between the dominant frequency of oscillation in the cells with the aggregate time-delay. We went on to show how asynchronous behaviour in a two-cell system corresponds to waves of gene expression in a chain of cells.

In a recent work \cite{keskin}, the gene expression noise in the PSM of the zebrafish was analysed. This was done by using smFISH microscopy techniques to count the numbers of discrete RNA molecules in individual cells. The statistical discrepancies between the gene expression in sets of cells which were supposedly synchronised was then evaluated (using well-known techniques \cite{elowitz, elowitz2}) and termed `expression noise'. It was found that this expression noise increased when mutations in both the \textit{DeltaC} and \textit{DeltaD} genes were introduced, reducing the efficacy of the Delta-Notch coupling. In our work, we have shown that increased Delta-Notch signalling strength increases the degree of synchronisation and the coherence of noisy oscillations, which in turn reduces the discrepancy between the cycles in coupled cells. This would appear to be very much in keeping with the aforementioned experimental findings.

On a more general note, intrinsic noise is often assumed to be a destabilising influence on cycles and to be the source of the discrepancy between the expression in two otherwise-equal cells. But, because of the complexity of the gene regulatory network and the nature of the coupling between cells in the PSM, the intrinsic noise can actually give rise to correlated sustained oscillations in neighbouring cells-- a behaviour one might normally associate with an extrinsic influence. This rather blurs the line between the what might be considered the signatures of intrinsic, extrinsic and `expression' noise in experimental data. As a result the utmost care must be taken to identify sources of correlation between cells, other than common external influence, if one is to truly discern the fingerprints of intrinsic noise in the data from those of extrinsic stochasticity.  

\section*{Author contributions}
JWB designed the study, contributed to discussions guiding the work, carried out mathematical calculations, performed simulations and wrote the manuscript. TG designed the study, contributed to discussions guiding the work and wrote the manuscript.

\section*{Funding statement}
JWB thanks the Engineering and Physical Sciences Research Council (EPSRC) for funding (PhD studentship, EP/N509565/1). TG acknowledges partial financial support from the Maria de Maeztu Program for Units of Excellence in R{\&}D (MDM-2017-0711). We declare no competing interests.

\pagebreak

\onecolumngrid

\begin{titlepage}
   \vspace*{\stretch{1.0}}
   \begin{center}
      \Large\textbf{Intrinsic noise, Delta-Notch signalling and delayed reactions promote sustained, coherent, synchronised oscillations in the presomitic mesoderm \\~\\
Supplemental Material}\\
   \end{center}
   \vspace*{\stretch{2.0}}
\end{titlepage}

\onecolumngrid
 \pagebreak
\setcounter{section}{0}		
\setcounter{page}{1}		
\setcounter{equation}{0}	
\setcounter{figure}{0}		
\setcounter{table}{0}		
\renewcommand{\thesection}{S\arabic{section}} 		
\renewcommand{\thepage}{S\arabic{page}} 			
\renewcommand{\theequation}{S\arabic{equation}}  	
\renewcommand{\thefigure}{S\arabic{figure}}  		
\renewcommand{\thetable}{S\arabic{table}} 
 
This supplement contains further details of our reduced model of the gene regulatory network as well as the calculations used to produce the results in the main paper. 

\section{Details of the model gene-regulatory system} \label{appendix:model}
In our analysis, we capture the intrinsic noise (which comes about due to the stochastic nature of the transcription/translation processes and finite particle numbers) using an individual-based model. In this model, we suppose that individual protein and mRNA molecules can be created or annihilated with certain probabilities per unit time, which may depend upon the various numbers of proteins/mRNAs currently in existence. There may also be delays between the initialisation of a creation/annihilation event and its completion. 

The dynamics, depicted in Fig. 1, are given more precisely by the following set of reactions,
\begin{align}
M_{d} &\stackrel[{\tau^{(p)d}}]{a_d}{\Longrightarrow} M_{d} + P_{d} , \nonumber \\
M_{h} &\stackrel[\tau^{(p)h}]{a_h}{\Longrightarrow} M_{h} + P_{h} , \nonumber \\
\emptyset &\stackrel[{\tau^{(m)d}}]{f^d}{\Longrightarrow} M_{d}  , \nonumber \\
\emptyset &\stackrel[\tau^{(m)h}]{f^h}{\Longrightarrow} M_{h} , \nonumber \\
P_d &\stackrel{b_d}{\longrightarrow}  \emptyset , \nonumber \\
P_h &\stackrel{b_h}{\longrightarrow}  \emptyset , \nonumber \\
M_{d} &\stackrel{c_d}{\longrightarrow}  \emptyset , \nonumber \\
M_{h} &\stackrel{c_h}{\longrightarrow} \emptyset , 
\label{reactscheme}
\end{align} 
where $M_d$ denotes a molecule of \textit{delta} mRNA and $P_h$ denotes a molecule of Hes/Her protein, etc. The single arrows $\stackrel{R}{\longrightarrow}$ indicate that the reaction occurs without delay with a per capita rate $R$. The double arrows $\stackrel[{\tau}]{R}{\Longrightarrow}$ indicate a delayed reaction with per capita rate $R$ and delay $\tau$. These equations are to be interpreted in the usual way using mass action kinetics \cite{erditoth}. For example, the first reaction is triggered with rate $a_d n^{(m)d}$, where $n^{(m)d}$ is the number of $M_d$-particles in the system. The effect of such a reaction is realised $\tau^{(p)d}$ units of time later, and results in the addition of a $P_d$-particle to the system.

The individual-based dynamics summarised by Eq.~(\ref{reactscheme}) can be approximated by a set of stochastic differential equations (SDEs), which are given in Section \ref{appendix:sdesmodel}.

The model parameters $a_{\rm h}, a_{\rm d}$, $b_{\rm h}, b_{\rm d}$ and $c_{\rm h}, c_{\rm d}$ are positive rate constants. The composite Hill functions  $f^h$ and $f^d$, encapsulating the activation/inhibition of mRNA production by the various protein concentrations are given by

\begin{align}
f^{\alpha}_j\left(\mathbf{n}_{t}\right) &= k_{\alpha}N\Bigg[r^{\alpha}_0 + r^{\alpha}_{\mathrm{d}} \frac{\frac{1}{2}\left(d^{(-)}\phi^{\mathrm{d}}_{j-1,t}+d^{(+)}\phi^{\mathrm{d}}_{j+1,t}\right)}{1+\frac{1}{2}\left(d^{(-)}\phi^{\mathrm{d}}_{j-1,t}+d^{(+)}\phi^{\mathrm{d}}_{j+1,t}\right)}+ r^{\alpha}_{\mathrm{h}} \frac{1}{1+ \phi^{\mathrm{h}}_{j,t}} +r^{\alpha}_{\mathrm{hd}} \frac{\frac{1}{2}\left(d^{(-)}\phi^{\mathrm{d}}_{j-1,t}+d^{(+)}\phi^{\mathrm{d}}_{j+1,t}\right)}{1+\frac{1}{2}\left(d^{(-)}\phi^{\mathrm{d}}_{j-1,t}+d^{(+)}\phi^{\mathrm{d}}_{j+1,t}\right)}\frac{1}{1+\phi^{\mathrm{h}}_{j,t} }\Bigg], \label{hillfuncts}
\end{align}

where $r_0^\alpha, r_{\rm d}^\alpha, r_{\rm h}^\alpha, r_{\rm hd}^\alpha$ and $k_\alpha$ are rate constants, $N$ is the system size, $\phi^{\alpha}_{j,t} = \frac{n^{\left(p\right)\alpha}_{j,t}}{n^{\left(p\right)\alpha}_0}$, and where  $d^{(\pm)}$ are positive constants such that $\frac{1}{2}\left(d^{(+)} + d^{(-)}\right)=1$; $n^{\left(p\right)\alpha}_0$ are reference values for the protein levels. Superscript or subscript indices $\alpha$ are placeholders, representing the cases $\alpha\in\{{\rm h,d}\}$. This reaction scheme is similar to the one used in \cite{lewis}. 

The different terms in the function $f^{\alpha}_j\left(\mathbf{n}_{t}\right)$ correspond to autonomous activation ($r^{\alpha}_0$), activation by \textit{delta} only ($r^{\alpha}_{\mathrm{d}}$), inhibition by \textit{hes/her} only ($r^{\alpha}_{\mathrm{h}}$) and mixed response to \textit{hes/her} and \textit{delta} $r^{\alpha}_{\mathrm{hd}}$. We constrain the parameters $r_0^\alpha, r_{\rm d}^\alpha, r_{\rm h}^\alpha, r_{\rm hd}^\alpha$ always to sum to unity i.e. $r^{\alpha}_0 + r^{\alpha}_{\mathrm{d}} + r^{\alpha}_{\mathrm{h}} + r^{\alpha}_{\mathrm{hd}} = 1 $. The expression inside the square brackets in Eq.~(\ref{hillfuncts}) can therefore range between $0$ and $1$ dynamically, depending on the concentrations of mRNA or protein and on the parameters $r_0^\alpha, r_{\rm d}^\alpha, r_{\rm h}^\alpha, r_{\rm hd}^\alpha$. As a result, the typical mRNA birth rate is characterised by $k_\alpha N$.

In the main text, we examine the change in the behaviour of coupled genetic oscillators as we vary the `coupling strength'. In order to isolate the effect of the Delta-Notch signalling from the \textit{hes/her} auto-repression, we always set $r^{\mathrm{h}}_0= r^{\mathrm{h}}_{\mathrm{d}}=0$ and vary $r^{\mathrm{h}}_{\mathrm{hd}}$ and $r^{\mathrm{h}}_{\mathrm{h}}$ such that $r^{\mathrm{h}}_{\mathrm{h}} + r^{\mathrm{h}}_{\mathrm{hd}} = 1 $. As $r^{\mathrm{h}}_{\mathrm{hd}}$ increases, so does the coupling strength but the role of \textit{hes/her} remains the same. For zero coupling $r^{\mathrm{h}}_{\mathrm{hd}}= 0$. For maximal coupling $r^{\mathrm{h}}_{\mathrm{hd}}= 1$. We keep the typical mRNA production rates $k_\alpha$ constant. 

In Sections III A, III B 1, III B 2 and III C, since we only consider a 2-cell model, \,\,\, $\frac{1}{2}\left(d^{(-)}\phi^{\mathrm{d}}_{j-1,t}+d^{(+)}\phi^{\mathrm{d}}_{j+1,t}\right) = \phi^{\mathrm{d}}_{j',t}$ where $j'=2$ if $j = 1$ and vice versa.

\medskip

\section{Quantification of stochastic fluctuations in systems with delays and non-local reaction rates}\label{appendix:derivationofsdes}
In this section, we derive the analytical results for the quantification of the stochastic fluctuations about the fixed point of a delay system with non-local reaction rates (Delta-Notch signalling). First, using a path integral approach, we derive expressions for the effective stochastic differential equations (SDEs) \cite{gardiner} which approximate the individual-based dynamics in the limit of large system-size $N$. We then use the linear-noise approximation to obtain an expression for the correlators of fluctuations about the deterministic fixed-point of the system. This procedure is similar to that used in \cite{brettgalla1,brettgalla2,barongalla}. These correlators are the basis for the theory results presented in the main text. Finally, we detail how this analysis can be used in conjunction with the model detailed in Section \ref{appendix:model} to produce the results in the main text.

\subsection{Path-integral approach}\label{appendix:ssexpansion}
We begin our analysis by defining a stochastic process in terms of the scaled variables $q^{\alpha}_{j,t} = n^{\alpha}_{j,t}/N$ (we use a more compact notation for the numbers of molecules $n^{\alpha}_{j,t}\equiv n^{\alpha}_{j}(t)$ than in the main text). Here, $N$ (defined in Section \ref{appendix:model}) characterises the typical number of particles per cell and is sometimes referred to as the system size \cite{vankampen}. To simplify matters, we discretise time in steps of length $\Delta$. The continuum limit is later recovered by taking $\Delta \to 0$. When reactions occur at a site $j$, particles may be created or annihilated immediately and/or at one other future time $t+\tau$. We define the number of reactions of type $r$ which occur in the time interval $t$ to $t+\Delta$ and have an associated delayed effect at $t+\tau$ by $k_{j,r,\tau,t}$. The number of particles of type $\alpha$ [where here $\alpha \in \{(p)d, (p)h, (m)d, (m)h \}$] which are immediately created/annihilated in such a reaction is denoted $\nu_r^\alpha$ and the number which are created/annihilated at the later time $t+\tau$ is denoted $w_{\tau,r}^\alpha$. 

The stochastic process can therefore be written as
\begin{align}
q^{\alpha}_{j,t+\Delta} - q^{\alpha}_{j,t} = \sum_{r,\tau} \frac{\nu_r^\alpha k_{j,r,\tau,t}}{N} + \sum_{r,\tau} \frac{ w_{\tau,r}^\alpha k_{j,r,\tau,t-\tau}}{N} .\label{stochprocess}
\end{align}
We wish to approximate this process with a set of stochastic differential equations. An elegant way to obtain this approximation is to start from a path-integral representation for the process, and then to perform an expansion in inverse powers of the system size (similar to that used by van Kampen \cite{vankampen}) within this representation. Using a path-integral approach, as opposed to a master equation, avoids the complications which arise due to the non-Markovian nature of the dynamics. 

We can write an expression for the probability of observing the system in a configuration $\mathbf{n}$ (which represents set of particle numbers of all types and locations) at time $t$ as follows

\begin{align}
&P\left(\mathbf{n},t\right) = \frac{1}{N}\int \left(\prod_{j,\alpha} \prod_{t' = 0}^{t}  dq^\alpha_{j, t'}\right) \delta\left[ \frac{1}{N}\mathbf{n} - \mathbf{q}\left(t\right)\right] \mathcal{P}\left(\{q^\alpha_{j, t'}\}\right), \label{pathintsimple}
\end{align}
where $\mathcal{P}\left(\{q^\alpha_{j, t'}\}\right)$ is the probability density of observing a particular trajectory (realisation of the system) $\{q^\alpha_{j, t'}\}$ and $\delta(\cdot)$ is the Dirac delta-function. Eq.~(\ref{pathintsimple}) is merely a statement that the probability of the system being in state $\mathbf{n}$ at time $t$ is equal to the sum of the probabilities of observing any one of a set of paths which lead to the system being in state $\mathbf{n}$ at time $t$. 

\medskip

Constraining the trajectory to obey Eq.~(\ref{stochprocess}) and then rewriting the Dirac delta functions as integrals of complex exponentials, one obtains
\begin{align}
&P\left(\mathbf{n},t\right)= \frac{1}{N}\sum_{\{ k\}}\int \left(\prod_{j,\alpha} \prod_{t' = 0}^{t}  dq^\alpha_{j, t'}\right) \delta\left[ \frac{1}{N}\mathbf{n} - \mathbf{q}\left(t\right)\right] \prod_{j,\alpha,\tau,t'}\delta\left(q^{\alpha}_{j, t'+\Delta} - q^{\alpha}_{j, t'} -  \sum_{r,\tau} \frac{\nu_r^\alpha k_{j,r,\tau,t'}}{N} - \sum_{r,\tau} \frac{ w_{\tau,r}^\alpha k_{j,r,\tau,t'-\tau}}{N} \right) P\left( \{k\}\right)\nonumber \\
&= \frac{1}{N}\sum_{\{ k\}}\int \left(\prod_{j,\alpha} \prod_{t' = 0}^{t}  \frac{dq^\alpha_{j, t'}dp^\alpha_{j,t'}}{2\pi}\right) \delta\left[\frac{1}{N}\mathbf{n} - \mathbf{q}\left(t\right)\right]  e^{i\sum_{j,\alpha,t'}p^{\alpha}_{j,t'}\left(q^{\alpha}_{j, t'+\Delta} - q^{\alpha}_{j, t'} -  \sum_{r,\tau} \frac{\nu_r^\alpha k_{j,r,\tau,t'}}{N} - \sum_{r,\tau} \frac{ w_{\tau,r}^\alpha k_{j,r,\tau,t'-\tau}}{N} \right)} P\left( \{k\}\right) , \label{pathint1}
\end{align}

where $P\left( \{k\}\right)$ is the probability of observing a particular set of creation/annihilation events $\{k\}$.

We presume that each reaction event is independent such that  
\begin{align}
P\left( \{k\}\right) = \prod_{j,r,\tau,t} P\left(k_{j,r,\tau,t} \right).
\end{align}
Once we specify the exact probability distributions for the variables $\{k\}$, we can evaluate the sums over $\{k\}$ in Eq.~(\ref{pathint1}). Let $W_{j,r,\tau,t}$ be the probability per unit time squared (or `rate') of a reaction of type $r$ occurring at position $j$ at a time between $t$ and $t+\Delta$ with associated delay between $\tau$ and $\tau+\Delta$. These rates may be dependent on the numbers of particles $\mathbf{n}$. We presume that the numbers of reactions triggered in the interval $t$ to $t+\Delta$, $k_{j,r,\tau,t}$, are Poisson random variables with mean $W_{j,r,\tau,t} \Delta^2$. This involves the approximation that the reaction rates do not change over the course of the small time step $\Delta$. This is similar to the approximation made for the so-called tau-leaping variant of the Gillespie algorithm \cite{tauleap}. Making these assumptions, the distributions of the $\{k\}$ are given by
\begin{align}
P\left(k_{j,r,\tau,t} \right) = \frac{\left(\Delta^2 W_{j,r,\tau,t}\right)^{k_{j,r,\tau,t}}}{k_{j,r,\tau,t}!}e^{-\Delta^2 W_{j,r,\tau,t}} .
\end{align}
We note that in our system, the local reaction rate $ W_{j,r,\tau,t}$ can depend upon the number of particles in the adjacent cells. The sums over $\{k\}$ in Eq.~(\ref{pathint1}) can be evaluated by observing that

\begin{align}
&\sum_{k_{j,r,\tau,t}}  \frac{\left(\Delta^2 W_{j,r,\tau,t}\right)^{k_{j,r,\tau,t}}}{k_{j,r,\tau,t}!}e^{-\Delta^2 W_{j,r,\tau,t}} e^{-i \sum_\alpha p^{\alpha}_{j,t}  \frac{\nu_r^\alpha k_{j,r,\tau,t}}{N} - i \sum_\alpha p^{\alpha}_{j,t+\tau} \frac{ w_{\tau,r}^\alpha k_{j,r,\tau,t}}{N} } \nonumber \\
&= \exp\left\{-\Delta^2 W_{j,r,\tau,t}\left[ 1- e^{-i \sum_\alpha \frac{\nu_r^\alpha p^{\alpha}_{j,t}+w_{\tau,r}^\alpha p^{\alpha}_{j,t+\tau}}{N}}\right] \right\}.
\end{align}
We finally arrive at the following expression for the probability distribution $P\left(\mathbf{n},t\right)$
\begin{align}
&P\left(\mathbf{n},t\right) =\frac{1}{N}\int \left(\prod_{j,\alpha} \prod_{t' = 0}^{t}  \frac{dq^\alpha_{j, t'}dp^\alpha_{j,t'}}{2\pi}\right) \delta\left[ \frac{1}{N}\mathbf{n} - \mathbf{q}\left(t\right)\right]  e^{i\sum_{j,\alpha,t'}p^{\alpha}_{j,t'}\left(q^{\alpha}_{j, t'+\Delta} - q^{\alpha}_{j, t'}  \right)}  \nonumber \\
&\times \exp\left\{-\sum_{j,r,\tau,t' }\Delta^2 W_{j,r,\tau,t'}\left[ 1- e^{-i\sum_\alpha \frac{\nu_r^\alpha p^{\alpha}_{j,t'}+w_{\tau,r}^\alpha p^{\alpha}_{j,t'+\tau}}{N}}\right] \right\}  . \label{pathint2}
\end{align}
Expanding the exponentials in Eq.~(\ref{pathint2}) in powers of $1/N$ and truncating the series at next-to-leading order, one obtains the following expression, 
\begin{align}
&P\left(\mathbf{n},t\right) = \frac{1}{N}\int \left(\prod_{j,\alpha} \prod_{t' = 0}^{t}  \frac{dq^\alpha_{j, t'}dp^\alpha_{j,t'}}{2\pi}\right) \delta\left[\frac{1}{N} \mathbf{n} - \mathbf{q}\left(t\right)\right]  e^{i\sum_{j,\alpha,t'}p^{\alpha}_{j,t'}\left(q^{\alpha}_{j, t'+\Delta} - q^{\alpha}_{j, t'}  \right)}  \nonumber \\
&\times \exp\Bigg\{-\sum_{j,r,\tau,t'}\Delta^2 W_{j,r,\tau,t'}\Bigg[  \frac{i}{N}\sum_\alpha\left(\nu_r^\alpha p^{\alpha}_{j,t'}+w_{\tau,r}^\alpha p^{\alpha}_{j,t'+\tau} \right) \nonumber \\
&+ \frac{1}{2 N^2} \sum_{\alpha, \alpha',j',t',\tau'}\delta_{\tau,\tau'}\delta_{j,j'}\delta_{t',t''}\left(\nu_r^\alpha p^{\alpha}_{j,t'}+w_{\tau,r}^\alpha p^{\alpha}_{j,t'+\tau} \right)\left(\nu_r^{\alpha'} p^{\alpha'}_{j',t''}+w_{\tau,r}^{\alpha'} p^{\alpha'}_{j',t''+\tau'}\right)\Bigg] \Bigg\}  , \label{pathint3}
\end{align}
where $\delta_{l,l'}$ is the Kronecker delta. Eq.~(\ref{pathint3}) is similar to the Martin-Siggia-Rose-Janssen-de Dominicis (MSRJD) functional integral \cite{altlandsimons}. This result can be compared with the path integral for an SDE of an appropriate form. The corresponding path integral expression for the general SDE 
\begin{align}
\frac{d n^\alpha_{j,t}}{dt} &= F^\alpha_{j,t}\left(\mathbf{n}_t \right) + \xi^\alpha_{j,t}, \nonumber \\
\langle \xi^\alpha_{j,t} \xi^{\alpha'}_{j',t'}\rangle &= \Sigma^{\alpha\alpha'}_{j,j',t,t'},
\end{align}
is given by \cite{brettgalla1, brettgalla2, barongalla}
\begin{align}
&P\left(\mathbf{n},t\right) =\frac{1}{N} \int \left(\prod_{j,\alpha} \prod_{t' = 0}^{t}  \frac{dq^\alpha_{j, t'}dp^\alpha_{j,t'}}{2\pi}\right) \delta\left[ \frac{1}{N}\mathbf{n} - \mathbf{q}\left(t\right)\right]  e^{i\sum_{j,\alpha,t'}p^{\alpha}_{j,t'}\left(q^{\alpha}_{j, t'+\Delta} - q^{\alpha}_{j, t'}  \right)}  \nonumber \\
&\times \exp\Bigg[-\frac{i \Delta}{N}\sum_{\alpha,j,t '} p^\alpha_{j,t'} F^\alpha_{j,t'}- \frac{\Delta^2}{2 N^2} \sum_{\alpha, \alpha',j,j',t',t''} p^\alpha_{j,t'}p^{\alpha'}_{j',t''} \Sigma^{\alpha\alpha'}_{j,j',t',t''} \Bigg] .
\end{align}
From Eq.~(\ref{pathint3}) one can then read off the following effective SDEs, which are good approximations of the dynamics for large but finite $N$, by taking the limit $\Delta \to 0$
\begin{align}
\frac{dn^\alpha_{j,t}}{dt} &= \sum_r W_{j,r,t}\nu^\alpha_{r}+\sum_r\int_0^\infty W_{j,r,\tau,t-\tau}w^\alpha_{r,\tau} d\tau + \xi^\alpha_{j,t}, \label{SDEs}
\end{align}
where $W_{j,r,t} = \int_0^\infty W_{j,r,\tau,t} d\tau $, and where the correlators of the stochastic noise variables are given by 
\begin{align}
\langle \xi^\alpha_{j,t} \xi^{\alpha'}_{j',t'}\rangle &= \delta\left(t-t'\right) \delta_{j,j'}\left[ \sum_{r} \nu_r^\alpha \nu_r^{\alpha'} W_{j,r,t} + \sum_{r} \int_0^\infty w^\alpha_{r,\tau}w^{\alpha'}_{r,\tau}W_{j,r,\tau,t-\tau}d\tau\right] \nonumber \\
+& \delta_{j,j'} \left[ \sum_r  W_{j,r,t-t',t}w^\alpha_{r,t-t'}\nu^{\alpha'}_{r} +\sum_r  W_{j,r,t'-t,t}w^{\alpha'}_{r,t'-t}\nu^\alpha_{r}  \right].\label{noisecorr}
\end{align}

\subsection{Linear-noise approximation and calculation of the power spectra and phase lags}\label{appendix:LNA}
We presume that the reaction rates can be decomposed as follows $W_{j,r,\tau,t} = W_{j,r,t}K_r\left(\tau\right)$. That is, the delay time $\tau$ is drawn from a distribution $K_r\left(\tau\right)$ which is independent of $j$, $t$ and $n^\alpha_{j,t}$. Eq.~(\ref{SDEs}) then becomes 
\begin{align}
\frac{dn^\alpha_{j,t}}{dt} &= \sum_r W_{j,r,t}\nu^\alpha_{r} \nonumber \\
&+\sum_r\int_0^\infty W_{j,r,t-\tau}K_r\left(\tau\right)w^\alpha_{r,\tau} d\tau + \xi^\alpha_{j,t}. \label{SDEs2}
\end{align}
If we consider small fluctuations about the fixed point of the deterministic system $\delta^\alpha_{j,t} = n^\alpha_{j,t} - \bar n^\alpha$ [as in Eq.~(1)], Eq.~(\ref{SDEs2}) may be approximated by 
\begin{align}
\frac{d\delta^\alpha_{j,t}}{dt} &= \sum_{j',\alpha'} J^{\alpha, \alpha'}_{j,j'}\delta^{\alpha'}_{j',t} +\sum_{j',\alpha'}\int_{-\infty}^\infty L^{\alpha,\alpha'}_{j,j',\tau}\delta^{\alpha'}_{j',t-\tau} d\tau + \xi^\alpha_{j,t}, \label{SDEslinear}
\end{align}
where 
\be
J^{\alpha,\alpha'}_{j,j'} = \left(\sum_r \frac{\partial W_{j,r,t}}{\partial n^{\alpha'}_{j',t}}\nu^\alpha_r\right)\Bigg\vert_{\left(\mathbf{n}_{j',t} = \bar{\mathbf{n}}\right)},
\ee
 and 
 \be
 L^{\alpha,\alpha'}_{j,j',\tau} =\left( \sum_r \frac{\partial W_{j,r,t-\tau}}{\partial n^{\alpha'}_{j',t-\tau}}K_r\left(\tau\right) \theta\left(\tau\right)w^\alpha_{r,\tau}\right) \Bigg\vert_{\left(\mathbf{n}_{j',t-\tau} = \bar{\mathbf{n}}\right)},
 \ee
where $\theta\left(\tau\right)$ is the Heaviside function. We note that for the systems studied in the main text, $J^{\alpha,\alpha'}_{j,j'}$ and $L^{\alpha,\alpha'}_{j,j',\tau}$ are non-zero for $j\neq j'$ due to the coupling of adjacent cells through Delta-Notch signalling. 

Crucially, as a part of this approximation, we now neglect fluctuations about the fixed point of the system in the evaluation of the correlators Eq.~(\ref{noisecorr}). That is, we evaluate $\langle \xi^\alpha_{j,t} \xi^{\alpha'}_{j',t'}\rangle $ at $n^{\alpha}_{j,t} = \bar{n}^\alpha$. As result, what was multiplicative noise is now treated as additive, simplifying the calculation. 

Carrying out a temporal Fourier transform in Eq.~(\ref{SDEslinear}), one obtains
\begin{align}
\sum_{\alpha',j'}\left[ i \omega \delta_{\alpha,\alpha'}\delta_{j,j'}-  J^{\alpha,\alpha'}_{j,j'} -\hat{ {L}}^{\alpha,\alpha'}_{j,j',\omega} \right] \hat{{\delta}}^{\alpha'}_{j',\omega} = \hat{{\xi}}^{\alpha}_{j,\omega} . \label{temporalonly}
\end{align}
Eq.~(\ref{temporalonly}) can be rewritten in matrix form as 
\begin{align}
\underline{\underline{M}}^{-1}_{\omega}\hat{{\underline{\delta}}}_{\omega} = \hat{{\underline{\xi}}}_{\omega}, \label{matrixm1}
\end{align}
where the different elements of the vector correspond to the different types of particle at the different cell sites.
Writing the correlation matrix of the noise variables as $\underline{\underline{\Sigma}}_{\omega} = \langle \hat{{\underline{\xi}}}_{\omega}\hat{{\underline{\xi}}}^\dagger_{\omega}\rangle$ one finally obtains the following result for the matrix of the correlators of the fluctuations
\begin{align}
\underline{\underline{C}}\left(\omega\right) \equiv \langle \hat{{\underline{\delta}}}_{\omega}\hat{{\underline{\delta}}}^\dagger_{\omega} \rangle = \underline{\underline{M}}_{\omega}\, \underline{\underline{\Sigma}}_{\omega}\, \underline{\underline{M}}^\dagger_{\omega} . \label{temporalonlycorr}
\end{align}
The diagonal elements of this matrix correspond to the power spectrum of fluctuations in Eq.~(2), i.e. $P^\alpha_j\left(\omega\right) = C^{\alpha,\alpha}_{j,j}\left(\omega\right)$. The off-diagonal elements allow one to calculate the phase lag as stated in Eq.~(3), that is $C^{\alpha,\alpha'}_{j,j'}\left(\omega\right) = \langle \hat\delta^\alpha_{j,\omega}\hat\delta^{\alpha'\star}_{j',\omega}\rangle$.

When the problem is translationally invariant, as is the case when the system parameters are the same in all cells, $J^{\alpha,\alpha'}_{j,j'}$ and $L^{\alpha,\alpha'}_{j,j',\tau}$ are functions of $j-j'$ only. One can then further simplify matters by carrying out a Fourier transform with respect to position as well. One obtains
\begin{align}
\sum_{\alpha'}\left[ i \omega \delta_{\alpha,\alpha'}- \tilde J^{\alpha,\alpha'}_{k} -\hat{\tilde {L}}^{\alpha,\alpha'}_{k,\omega} \right] \hat{\tilde{\delta}}^{\alpha'}_{k,\omega} = \hat{\tilde{\xi}}^{\alpha}_{k,\omega}, \label{fouriercomponents}
\end{align}
and from this
\begin{align}
\underline{\underline{M}}^{-1}_{k,\omega}\hat{\tilde{\underline{\delta}}}_{k,\omega} = \hat{\tilde{\underline{\xi}}}_{k,\omega}, \label{matrixm2}
\end{align}
where the different elements of the vector now correspond only to the different species. We then arrive at a similar expression to Eq.~(\ref{temporalonlycorr}), but where the matrix dimension is reduced by a factor of $L$ (the number of cells)
\begin{align}
\underline{\underline{C}}\left(k,\omega\right) \equiv \langle \hat{\tilde{\underline{\delta}}}_{k,\omega}\hat{\tilde{\underline{\delta}}}^\dagger_{k,\omega} \rangle = \underline{\underline{M}}_{k,\omega}\underline{\underline{\Sigma}}_{k,\omega} \underline{\underline{M}}^\dagger_{\omega} . \label{bothcorr}
\end{align}
Here, the diagonal elements correspond to the power spectra in Eq.~(6) and the off-diagonal elements allow one to calculate the phase lag using Eq.~(3) of the main text.

\subsection{Application to the model gene regulatory network}\label{appendix:sdesmodel}

Using Eqs.~(\ref{SDEs}) and (\ref{noisecorr}), the individual-based system given by Eq.~(\ref{reactscheme}) can be approximated by stochastic differential equations of the form
\begin{align}
\frac{d n^{\mathrm{(p)h}}_{j,t}}{dt} &=  a_{\mathrm{h}} n^{\mathrm{(m)h}}_{j,t-\tau^{\mathrm{(p)h}}} - b_{\mathrm{h}}n^{\mathrm{(p)h}}_{j,t} + \xi^{\mathrm{(p)h}}_{j,t} , \nonumber \\
\frac{d n^{\mathrm{(p)d}}_{j,t}}{dt} &=  a_{\mathrm{d}} n^{\mathrm{(m)d}}_{j,t-\tau^{\mathrm{(p)d}}} - b_{\mathrm{d}}n^{\mathrm{(p)d}}_{j,t} + \xi^{\mathrm{(p)d}}_{j,t} , \nonumber \\
\frac{d n^{\mathrm{(m)h}}_{j,t}}{dt} &=  f^{\mathrm{h}}_j\left(\mathbf{n}_{t-\tau^{\mathrm{(m)h}}}\right) - c_{\mathrm{h}}n^{\mathrm{(m)h}}_{j,t} + \xi^{\mathrm{(m)h}}_{j,t} , \nonumber \\
\frac{d n^{\mathrm{(m)d}}_{j,t}}{dt} &=  f^{\mathrm{d}}_j\left(\mathbf{n}_{t-\tau^{\mathrm{(m)d}}}\right) - c_{\mathrm{d}}n^{\mathrm{(m)d}}_{j,t} + \xi^{\mathrm{(m)d}}_{j,t}. \label{diffeqs2}
\end{align}
The quantities involved in these equations are given in Section \ref{appendix:model}, with the exception of the Gaussian noise terms $\xi^\alpha_{j,t}$, which encapsulate the intrinsic noise due to the stochastic and individual-based nature of the gene regulatory system. The deterministic trajectories $\bar{n}^\alpha_{j,t}$, examples of which are given in Figs. 2 and 3, are found by setting the noise terms $\xi^\alpha_{j,t}$ in Eq.~(\ref{diffeqs2}) to zero and numerically integrating the resulting ordinary differential equations. Setting the noise terms to zero effectively approximates the system as being infinitely large, so that any fluctuations are negligible in comparison to the mean particle numbers. This is not appropriate in cases in which intrinsic noise significantly affects the dynamics (as is the case for the examples we look at). This is exemplified by the stark disagreement between the deterministic and individual-based simulations in Figs. 2 and 3. 

When the deterministic (infinite) system has reached a fixed point, the noise terms $\xi^\alpha_{j,t}$ in Eq.~(\ref{diffeqs2}) can be taken to have the following correlators within the linear-noise approximation,
\begin{align}
\langle  \xi^{\mathrm{(p)h}}_{j,t} \xi^{\mathrm{(p)h}}_{j',t'}  \rangle &= \delta\left(t-t'\right) \delta_{j,j'} \left[a_{\mathrm{h}} \bar n^{\mathrm{(m)h}} + b_{\mathrm{h}}\bar n^{\mathrm{(p)h}} \right],\nonumber \\
\langle \xi^{\mathrm{(p)d}}_{j,t} \xi^{\mathrm{(p)d}}_{j',t'} \rangle &= \delta\left(t-t'\right) \delta_{j,j'} \left[a_{\mathrm{d}} \bar n^{\mathrm{(m)d}} + b_{\mathrm{d}}\bar n^{\mathrm{(p)d}} \right],\nonumber \\
\langle  \xi^{\mathrm{(m)h}}_{j,t} \xi^{\mathrm{(m)h}}_{j',t'} \rangle &= \delta\left(t-t'\right) \delta_{j,j'} \left[f_{\mathrm{h}}\left(\mathbf{\bar n}\right) + c_{\mathrm{h}}\bar n^{\mathrm{(m)h}} \right],\nonumber \\
\langle  \xi^{\mathrm{(m)d}}_{j,t}\xi^{\mathrm{(m)d}}_{j',t'} \rangle &= \delta\left(t-t'\right) \delta_{j,j'} \left[f_{\mathrm{d}}\left(\mathbf{\bar n}\right) + c_{\mathrm{d}}\bar n^{\mathrm{(m)d}} \right], \label{correlatorsgenereg}
\end{align}
where barred quantities are evaluated at the deterministic fixed point (i.e. the solution to Eqs.~(\ref{diffeqs2}) with the noise terms and the time derivatives set to zero). All inter-species cross-correlators are zero, due to the fact that no one reaction gives rise to the production/annihilation of two different species of particle (see Eq.~\ref{reactscheme}).

In order to evaluate the power spectrum of fluctuations or to find the phase lag, one first finds the matrix $\underline{\underline{M}}_{\omega}$ or $\underline{\underline{M}}_{k,\omega}$, defined in Eqs.~(\ref{matrixm1}) or (\ref{matrixm2}) respectively, by performing the linear-noise approximation on Eq.~(\ref{diffeqs2}), as detailed in Section \ref{appendix:LNA}. Using the correlators in Eq.~(\ref{correlatorsgenereg}), one is then able to evaluate the matrix of correlators $\underline{\underline{C}}\left(\omega\right)$ or $\underline{\underline{C}}\left(k,\omega\right)$, defined in Eqs.~(\ref{temporalonlycorr}) or (\ref{bothcorr}) respectively, which contains all the information one needs to find the power spectra of fluctuations and/or the phase lag (as is also described in Section~\ref{appendix:LNA}).

As an example, we provide expressions that allow one to calculate the matrix $\underline{\underline{M}}_{k,\omega}$ in the case that the system is spatially homogeneous. In Eq.~(\ref{fouriercomponents}), the matrix $\tilde J_k^{\alpha \alpha'}$ is given by
\begin{align}
\underline{\underline{\tilde J_k}} = \begin{bmatrix}
    -b_h & 0 & 0  & 0\\
    0 & -b_d & 0 &  0 \\
    0 & 0 & -c_h  &0 \\
     0 & 0 & 0  & -c_d 
\end{bmatrix} ,
\end{align}
and the matrix $\hat{\tilde{L}}^{\alpha \alpha'}_{k,\omega}$ is given by
\begin{align}
\underline{\underline{\hat{\tilde{L}}_{k,\omega}}} = \begin{bmatrix}
    0 & 0 & a_h e^{i \omega \tau^{\mathrm{(p)h}}}  & 0\\
    0 & 0 & 0 &  a_d e^{i \omega \tau^{\mathrm{(p)d}}} \\
    \left(f^{\mathrm{h}}_{h} e^{i \omega \tau^{\mathrm{(m)h}}} \right) & \left(f^{\mathrm{h}}_{d} e^{i \omega \tau^{\mathrm{(m)h}}} \frac{1}{2}(d^{(+)}e^{-ik} + d^{(-)}e^{ik})\right) & 0  &0 \\
    \left(f^{\mathrm{d}}_{h} e^{i \omega \tau^{\mathrm{(m)d}}}\right) & \left(f^{\mathrm{d}}_{d} e^{i \omega \tau^{\mathrm{(m)d}}} \frac{1}{2}(d^{(+)}e^{-ik} + d^{(-)}e^{ik})\right) & 0  &0 
\end{bmatrix} ,
\end{align}
where the entries of the matrix are in the order $\alpha =\{ \mathrm{(p)h}, \,\, \mathrm{(p)d}, \,\, \mathrm{(m)h}, \,\, \mathrm{(m)d}\}$ and where we have
\begin{align}
f^{\mathrm{\alpha}}_{h} &= -k_\alpha N \frac{n_0^{(p)h}}{(n_0^{(p)h} + \bar n^{(p)h})^2} \left[ r^\alpha_{\mathrm{h}} + r^\alpha_{\mathrm{hd}} \frac{\bar n^{(p)d}}{n_0^{(p)d}+\bar n^{(p)d} } \right] ,  \nonumber \\
f^{\mathrm{\alpha}}_{d} &= k_\alpha N \frac{n_0^{(p)d}}{(n_0^{(p)d}+\bar n^{(p)d} )^2} \left[ r^\alpha_{\mathrm{d}} + r^\alpha_{\mathrm{hd}} \frac{n_0^{(p)h}}{n_0^{(p)h} + \bar n^{(p)h}} \right].
\end{align}

\section{Linear stability analysis}\label{appendix:lsa}
We have derived expressions for the power spectrum of fluctuations in the individual-based system about the deterministic fixed point. Such an analysis presumes the stability of this fixed point. To determine whether or not such a stable fixed point exists for a particular parameter set, we perform a linear stability analysis of the deterministic system. 

We begin with the linearised expression for the time-evolution of small deviations about the deterministic fixed point Eq.~(\ref{SDEslinear}), but with the noise term removed 
\begin{align}
\frac{d\delta^\alpha_{j,t}}{dt} &= \sum_{j',\alpha'} J^{\alpha, \alpha'}_{j,j'}\delta^{\alpha'}_{j',t} +\sum_{j',\alpha'}\int_0^\infty L^{\alpha,\alpha'}_{j,j',\tau}\delta^{\alpha'}_{j',t-\tau} d\tau . \label{SDEslinearnonoise}
\end{align}
For simplicity, we now assume that the delay kernels are all Dirac-delta functions [i.e. $K_r\left(\tau\right)= \delta\left(\tau- \tau_r\right)$] such that 
\begin{align}
\sum_{j',\alpha'}\int_0^\infty L^{\alpha,\alpha'}_{j,j',\tau}\delta^{\alpha'}_{j',t-\tau} d\tau = \sum_{j',\alpha',r}H^{\alpha,\alpha'}_{j,j',r}\delta^{\alpha'}_{j',t-\tau_{r}},
\end{align}
where 
\be
H^{\alpha,\alpha'}_{j,j',r}=\left( \frac{\partial W_{j,r,t-\tau}}{\partial n^{\alpha'}_{j',t-\tau}} w^\alpha_{r}\right) \Bigg\vert_{\left(\mathbf{n}_{j',t-\tau} = \bar{\mathbf{n}}\right)}.
\ee
We have further supposed that the number of particles produced by a delayed reaction $w^\alpha_{r}$ is independent of the delay time $\tau_r$. 

We then suppose that the deviation away from the fixed point evolves in an exponential way. That is we propose the ansatz 
\begin{align}
\delta_{j,t}^\alpha = \delta_{j}^{\alpha (0)} e^{\lambda t} . \label{expansatz}
\end{align} 
Upon substitution into Eq.~(\ref{SDEslinearnonoise}), one obtains
\begin{align}
 \sum_{j',\alpha'}\left[\lambda\delta_{j,j'}\delta_{\alpha,\alpha'} - J^{\alpha, \alpha'}_{j,j'} -  \sum_r H^{\alpha,\alpha'}_{j,j',r}e^{\lambda \tau_r}\right]\delta_{j'}^{\alpha' (0)} = 0. \label{withindices}
\end{align}
We note that if we had not assumed a delta function for the delay kernel $K_r\left(\tau\right)$, the exponential factors in Eq.~(\ref{eigenequation}) would instead be replaced by $\mathcal{L}_\tau\left\{K_r\left(\tau\right)\right\}\left(-\lambda\right)$, the Laplace transform of the delay kernel evaluated at $-\lambda$. Eq.~(\ref{withindices}) can be rewritten more succinctly as the following matrix equation
\begin{align}
\left[\lambda\underline{\underline{\mathbb{1}}} - \underline{\underline{J}} -  \sum_r \underline{\underline{H}}_{r}e^{\lambda \tau_r}\right]\underline{\delta}^{(0)} = 0.\label{eigenequation}
\end{align}
In order for the solutions to this equation to be non-trivial, the determinant of the object multiplying $\underline{\delta}^{(0)}$ in Eq.~(\ref{eigenequation}) must be equal to zero. This gives rise to an `eigenvalue' equation for $\lambda$. Due to the exponential terms involving $\lambda$ which arise from the delays, this equation is not analytically tractable and must be solved numerically. 

Typically, one finds many possible solutions $\lambda$ to the eigenvalue equation for any one set of system parameters, and thus the full solution to Eq.~(\ref{SDEslinearnonoise}) is a linear combination of solutions of the form in Eq.~(\ref{expansatz}). If any one of the eigenvalues $\lambda$ has a positive real part the fixed point is unstable. Else, we say that it is (linearly) stable. In order to produce the dotted lines in Figs. 5 and 6, one finds the eigenvalues $\lambda$ as a function of the system parameters. The instability line divides regions in parameter space for which all the eigenvalues have negative real parts from those regions for which some eigenvalues have positive real parts. The imaginary part of the least stable eigenvalue for each parameter set is plotted as a green line in Figs. 5a and 6a. 

Once one has found the eigenvalues $\lambda$, one can then also go on to solve (numerically) for the eigenvectors $\underline{\delta}^{(0)}$. These yield information about the relative amplitudes and phases of the various components. The complex phase difference between elements of the eigenvector that correspond to the same species in opposite cells is shown as a green line in Figs. 5b and 6b. The inter-cell phase difference turns out to be the same for all particle types.

\end{document}